\documentclass{pasj00}
\draft

\begin{document}
\SetRunningHead{S. Kato}
               {}
\Received{2013/0/00}
\Accepted{2013/0/00}

\title{Resonant Excitation of Disk Oscillations in Deformed Disks VII: Stability Criterion in MHD Systems} 

\author{Shoji \textsc{Kato}}%
\affil{2-2-2 Shikanodai-Nishi, Ikoma-shi, Nara 630-0114}
\email{kato.shoji@gmail.com; kato@kusastro.kyoto-u.ac.jp}


%

\KeyWords{accretion, accretion disks --- magnetic fields --- oscillations   
    --- resonance --- stability} 

\maketitle

\begin{abstract}
In a disk with an oscillatory deformation from an axisymmetric state 
with frequency $\omega_{\rm D}$ and azimuthal wavenumber $m_{\rm D}$,
a set of two normal mode oscillations with frequency and azimuthal wavenumber being ($\omega_1$, $m_1$)
and ($\omega_2$, $m_2$) resonantly couple through the disk deformation,
when the resonant conditions ($\omega_1+\omega_2+\omega_{\rm D}=0$ and $m_1+m_2+m_{\rm D}=0$)
are satisfied.
In the case of hydrodynamical disks, the resonance amplifies the set of the oscillations if
$(E_1/\omega_1)(E_2/\omega_2)>0$ (Kato 2013b), where $E_1$ and $E_2$ are wave energies of the two
oscillations with $\omega_1$ and $\omega_2$, respectively.
In this paper we show that this instability criterion is still valid even when the oscillations are ideal MHD ones
in magnetized disks, if the displacements associated with the oscillations vanish on the boundary of the system.
\end{abstract}


\section{Introduction}

Energy consideration is primary significance in understanding various hydrodynamic and hydromagnetic
instabilities.
In a static system, energy associated with perturbations is always positive.
In such systems stability can be examined by using the energy principle developed by Bernstein et al.
(1958).
In a stationary system with a shear flow, however, energy associated with perturbations is not always positive and
can become negative.
Well-known examples of negative wave-energy perturbations are waves in such rotating systems as galactic and accretion disks,
where oscillations outside the radius of the corotation resonance have positive energy, while those inside have negative
energy.
It is known that in such systems two-wave coupling with opposite signs of wave energy can lead to instability.
For examples, spiral density waves in galactic disks are sustained by wave amplification (over-reflection) 
at the corotation resonance (Lin \& Lau 1980).
Papaloizou and Pringle instability (Drury 1985, Papaloizou \& Pringle 1984) is also related to interaction between 
positive- and  negative-energy waves through corotation radius.
Amplification of p-mode oscillations at the corotation resonance in accretion disks  (Lai \& Tsang 2009, see also
Fu and Lai 2011) also belongs to the instability of the same category.

A classical example which can be interpreted in terms of positive- and negative-wave coupling is Kelvin-Helmholtz instability
(Cairns 1978).
Especially, Khalzov et al. (2007) suggest that coupling of waves with positive and negative energies is a universal mechanism for 
MHD instabilities of flowing media.

Importance of energy consideration is also recognized in a wave-wave resonant instability in deformed  
hydrodynamical disks (Kato et al. 2011 and Kato 2013b).\footnote{
This wave-wave resonant instability is essentially a generalization of Lubow's tidal instability (1991)
from a view point emphasizing wave phenomena.
} 
In these studies, however, we find that what is directly related to stability criterion is not the sign of
wave energy, $E$, itself, but the sign of $E/\omega$, where $\omega$ is the frequency of oscillations.
That is, let us consider a disk deformed from an axisymmetric state, 
where the deformation has frequency $\omega_{\rm D}$ and azimuthal wavenumber $m_{\rm D}$.
On such deformed disks, two small-amplitude normal mode oscillations are superposed.
The set of frequency and azimuthal wavenumber, ($\omega$, $m$), of these two normal mode oscillations are 
($\omega_1$, $m_1$) and ($\omega_2$, $m_2$).
If
\begin{equation}
      \omega_1+\omega_2+\omega_{\rm D}=0, \qquad m_1+m_2+m_{\rm D}=0
\end{equation}
are realized, the two oscillations with $(\omega_1$, $m_1$) and ($\omega_2$, $m_2$) resonantly couple each other 
through the disk deformation.\footnote{
Readers may think that an additional resonant condition is necessary: a relation among vertical 
node numbers of oscillations.
It is, however, unnecessary to consider it here, since if there is an additional condition and if it is not satisfied,
the coupling term $W$ [see equation (\ref{3.3})] vanishes automatically.
}
The instability condition is found to be (Kato 2013b)
\begin{equation}
        \frac{E_1}{\omega_1}\frac{E_2}{\omega_2}>0.
\label{condition}
\end{equation}
In particular cases where $\omega_{\rm D}$ is sufficiently low, the resonant condition $\omega_1+\omega_2+\omega_{\rm D}=0$
is realized for $\omega_1\omega_2<0$.
Then, the instability condition (\ref{condition}) is reduced to $E_1E_2<0$.
Except for such cases, however, the instability condition is generally given by $(E_1/\omega_1)(E_2/\omega_2)>0$.

The purpose of this paper 
is to show that the above instability condition, $(E_1/\omega_1)(E_2/\omega_2)>0$, 
can be extended even to the cases of ideal MHD disks, under the condition that displacements associated with the oscillations 
vanish on the boundary of the system. 
Except that the disks are subject to magnetic fields, the procedures of analyses in this paper are quite parallel 
to those by Kato (2013b, referred to paper I).
Hence, the parts almost parallel to those in paper I will be described only briefly. 

\section{Linearized Hydromagnetic Equations and Normal Mode Oscillations}

The unperturbed disk is steady and axisymmetric.
In the Lagrangian formulation, hydromagnetic perturbations 
superposed on the unperturbed disks can be described by extending the hydrodynamical formulation by Lynden-Bell \& Ostriker (1967) 
to hydromagnetic cases as
\begin{equation}
       \frac{D_0^2\mbox{\boldmath $\xi$}}{Dt^2}=\Delta\biggr(-\nabla\psi-\frac{1}{\rho}\nabla p
             +\frac{1}{\rho}{\rm curl}\mbox{\boldmath $B$}\times \mbox{\boldmath $B$}\biggr),
\label{2.1}
\end{equation}
where $\mbox{\boldmath $\xi$}(\mbox{\boldmath $r$}, t)$ is a displacement vector associated with the perturbations, 
and $D_0/Dt$ is the time derivative along an unperturbed flow, $\mbox{\boldmath $u$}_0(\mbox{\boldmath $r$})$, and is related to 
the Eulerian time derivative, $\partial/\partial t$,  by
\begin{equation}
      \frac{D_0}{Dt}=\frac{\partial}{\partial t}+\mbox{\boldmath $u$}_0\cdot\nabla.
\label{2.2}
\end{equation}
In equation (\ref{2.1}) $\Delta(X)$ represents the Lagrangian variation of $X$, and $\psi$ is the gravitational
potential.
Other notations in equation (\ref{2.1}) have their usual meanings.

In the case where the perturbations have small amplitude and are non-dissipative, 
equation (\ref{2.1}) is written as 
\begin{equation}
     \rho_0\frac{\partial^2\mbox{\boldmath $\xi$}}{\partial t^2}
       +2\rho_0(\mbox{\boldmath $u$}_0\cdot\nabla)\frac{\partial\mbox{\boldmath $\xi$}}{\partial t}
       +\mbox{\boldmath $L$}(\mbox{\boldmath $\xi$})=0,
\label{2.3}
\end{equation}
where $\mbox{\boldmath $L$}(\mbox{\boldmath $\xi$})$ consists of hydrodynamic and hydromagnetic parts as
\begin{equation}
    \mbox{\boldmath $L$}(\mbox{\boldmath $\xi$})=\mbox{\boldmath $L$}^{\rm G}(\mbox{\boldmath $\xi$})
                                               + \mbox{\boldmath $L$}^{\rm B}(\mbox{\boldmath $\xi$}).
\label{L}
\end{equation}
Detailed expressions for $\mbox{\boldmath $L$}^{\rm G}(\mbox{\boldmath $\xi$})$ and 
$\mbox{\boldmath $L$}^{\rm B}(\mbox{\boldmath $\xi$})$ are unnecessary here.
What we need here is that both of them  are Hermitian in the following sense: 
\begin{equation}
      \int \mbox{\boldmath $\eta$}\cdot\mbox{\boldmath $L$}^{\rm G}(\mbox{\boldmath $\xi$})dV
      =\int \mbox{\boldmath $\xi$}\cdot\mbox{\boldmath $L$}^{\rm G}(\mbox{\boldmath $\eta$})dV,
         \qquad
       \int \mbox{\boldmath $\eta$}\cdot\mbox{\boldmath $L$}^{\rm B}(\mbox{\boldmath $\xi$})dV
       =\int \mbox{\boldmath $\xi$}\cdot\mbox{\boldmath $L$}^{\rm B}(\mbox{\boldmath $\eta$})dV,  
\label{Hermitian}
\end{equation}
where $\mbox{\boldmath $\xi$}$ and $\mbox{\boldmath $\eta$}$ are any non-singular functions of $\mbox{\boldmath $r$}$ 
defined in the unperturbed volume of the disk and having continous first and second derivatives everwhere.
The integration is performed over the whole volume of the system, assuming that the surface integrals vanish.
The Hermitian of the hydrodynamical part, $\mbox{\boldmath $L$}^{\rm G}(\mbox{\boldmath $\xi$})$, is shown by 
Lynden-Bell \& Ostriker (1967) and that of the hydromagnetic part, $\mbox{\boldmath $L$}^{\rm B}(\mbox{\boldmath $\xi$})$, for example,
by Bernstein et al. (1958) and Khalzov et al. (2008). 
It is noted that $\mbox{\boldmath $L$}(\mbox{\boldmath $\xi$})$ is Hermitian even when the perturbations are self-gravitating,
but we consider hereafter only the cases of non-selfgravitating perturbations.
  
We now consider three normal mode oscillations satisfying the linearized equation (\ref{2.3}). 
The set of eigen-frequency and azimuthal wavenumber of these oscillations are denoted by
($\omega_1$, $m_1$), ($\omega_2$, $m_2$), and ($\omega_3$, $m_3$).
The displacement vectors, $\mbox{\boldmath $\xi$}_{\rm i}(\mbox{\boldmath $r$}, t)$, associated with these 
oscillations are expressed as 
\begin{equation}
    \mbox{\boldmath $\xi$}_{\rm i}(\mbox{\boldmath $r$}, t)=\Re\biggr[\hat{\mbox{\boldmath $\xi$}}_{\rm i}
            (\mbox{\boldmath $r$}){\rm exp}(i\omega_{\rm i}t)\biggr]
      =\Re\biggr[\breve{\mbox{\boldmath $\xi$}}_{\rm i}{\rm exp}[i(\omega_{\rm i} t-m_{\rm i}\varphi)]\biggr] \qquad ({\rm i}=1, 2, 3),
\label{2.2'}
\end{equation}
where $\Re$ denotes the real part, and $\varphi$ is the azimuthal coordinate of the cylindrical coordinates ($r$, $\varphi$, $z$) 
whose center is at the disk center and the $z$-axis is the rotating axis of the disk.
We now assume that the following resonant conditions among the three oscillations are present:
\begin{equation}
    \omega_1+\omega_2+\omega_3=\Delta\omega, \quad {\rm and}\quad m_1+m_2+m_3=0,
\label{2.3'}
\end{equation}
where $m_{\rm i}$'s (${\rm i}=1,2,3$) are integers.
In order to include in our formulation the cases where three frequencies are slightly deviated from the exact resonant
condition, $\Delta\omega$ is introduced in the first relation of equation (\ref{2.3'}), where $\vert\Delta\omega\vert$ 
is assumed to be much smaller than the absolute values of $\omega_1$ and $\omega_2$.\footnote{
The cases where $\omega_3$ (which is denoted later by $\omega_{\rm D})=0$ can be included in our analyses.
}

Our purpose in this paper is to examine how the resonant interactions among three oscillations change their amplitudes.
Our main interest, however, is the case where among three oscillations, the ($\omega_3$, $m_3$) oscillation has 
a particulat position in the sense that it has a larger amplitude compared with the other two and its amplitude 
variation due to resonant interactions with other modes can be practically neglected or it is maintained at a fixed 
amplitude by such an external force as tidal one.
In other words, we consider resonant interactions between two oscillations with ($\omega_1$, $m_1$) and ($\omega_2$, $m_2$)
in a deformed disk with ($\omega_3$, $m_3$).
In order to emphasize this situation, $\mbox{\boldmath $\xi$}_{\rm 3}(\mbox{\boldmath $r$}, t)$ and ($\omega_3$, $m_3$)
are denoted  hereafter $\mbox{\boldmath $\xi$}_{\rm D}(\mbox{\boldmath $r$}, t)$ and ($\omega_{\rm D}$, $m_{\rm D}$),
respectively.

Before examining the resonant interaction, we introduce the wave energy of oscillations, $E_{\rm i}$ (i $=$ 1, 2), 
defied by (e.g., Kato 2001)
\begin{equation}
      E_{\rm i}=\frac{1}{2}\omega_{\rm i}\biggr[\omega_1\langle\rho_0\hat{\mbox{\boldmath $\xi$}}_{\rm i}^*
                  \hat{\mbox{\boldmath $\xi$}}_{\rm 1}\rangle
          -i\langle\rho_0\hat{\mbox{\boldmath $\xi$}}_{\rm i}^*(\mbox{\boldmath $u$}_0\cdot\nabla)
                  \hat{\mbox{\boldmath $\xi$}}_{\rm i}\rangle\biggr]
          =\frac{1}{4}\omega_{\rm i}\biggr[\omega_{\rm i}\langle\rho_0\hat{\mbox{\boldmath $\xi$}}_{\rm i}^*
                  \hat{\mbox{\boldmath $\xi$}}_{\rm i}\rangle
           +\langle\hat{\mbox{\boldmath $\xi$}}_{\rm i}^*\cdot\mbox{\boldmath $L$}(\hat{\mbox{\boldmath $\xi$}}_{\rm i})\rangle\biggr],
\label{2.14}
\end{equation}
where the asterisk denotes the complex conjugate, and $\langle X \rangle$ is the volume integration of $X$.
In the framework of linear theory, the wave energy (\ref{2.14}) of each oscillation is conserved, 
i.e., $\partial E_{\rm i}/\partial t= 0$,
which is derived easily by using the facts that the operator $\mbox{\boldmath $L$}(\mbox{\boldmath $\xi$}_i)$
is Hermitian and $\mbox{\boldmath $\xi$}_{\rm i}$ follows the linear wave equation (\ref{2.3}).
It is instructive to note that the wave energy is expressed in the case of the oscillations in geometrically thin non-magnetized
disks as (e.g., see Kato 2001)
\begin{equation}
    E_{\rm i}= \frac{\omega_1}{2}\biggr\langle(\omega_{\rm i}-m_{\rm i}\Omega)\rho_0(\hat{\xi}_{{\rm i},r}^*\hat{\xi}_{{\rm i},r}
              +\hat{\xi}_{{\rm i},z}^*\hat{\xi}_{{\rm i},z})\biggr\rangle \qquad ({\rm i}=1,2),
\label{energy}
\end{equation}
where $\Omega(r)$ is the angular velocity of disk rotation.
It is noticed that the sign of wave energy depends on which side of the corotation radius
($\omega_{\rm i}=m_{\rm i}\Omega$) the oscillation is.

\section{Formulation of Coupling Processes and Commutability of Coupling Terms}
 
Now, we consider two normal modes of oscillations, $\mbox{\boldmath $\xi$}_1$ and $\mbox{\boldmath $\xi$}_2$,
with ($\omega_1$, $m_1$) and ($\omega_2$, $m_2$), respectively.
In the linear stage, the perturbation, $\mbox{\boldmath $\xi$}$, imposed on a deformed disk is simply the sum 
of these two oscillations:
\begin{eqnarray}
        \mbox{\boldmath $\xi$}(\mbox{\boldmath $r$}, t) 
          = &&A_1 \mbox{\boldmath $\xi$}_1(\mbox{\boldmath $r$}, t)
            + A_2 \mbox{\boldmath $\xi$}_2(\mbox{\boldmath $r$}, t)      
                                  \nonumber    \\
          =&&\Re\biggr[A_1\breve{\mbox{\boldmath $\xi$}}_1{\rm exp}[i(\omega_1t-m_1\varphi)]
                      +A_2\breve{\mbox{\boldmath $\xi$}}_2{\rm exp}[i(\omega_2t-m_2\varphi)]\biggr],
\label{2.6}
\end{eqnarray}
where $A_1$ and $A_2$ are amplitudes and are arbitrary constants.
In the quasi-nonlinear stage, the oscillations  are coupled through disk deformation, 
$\mbox{\boldmath $\xi$}_{\rm D}$, so that they satisfy a quasi-nonlinear wave equation. 
The non-linear wave equation is
\begin{equation}
    \rho_0\frac{\partial^2\mbox{\boldmath $\xi$}}{\partial t^2}
       +2\rho_0(\mbox{\boldmath $u$}_0\cdot\nabla)\frac{\partial\mbox{\boldmath $\xi$}}{\partial t}
       +\mbox{\boldmath $L$}(\mbox{\boldmath $\xi$})
    = \mbox{\boldmath $C$}(\mbox{\boldmath $\xi$}, \mbox{\boldmath $\xi$}_{\rm D}),
\label{2.7}
\end{equation}
where $\mbox{\boldmath $C$}$ is the quasi-nonlinear coupling terms and consists of hydrodynamic and hydromagnetic terms:
\begin{equation}
         \mbox{\boldmath $C$}(\mbox{\boldmath $\xi$}, \mbox{\boldmath $\xi$}_{\rm D})
         =\mbox{\boldmath $C$}^{\rm G}(\mbox{\boldmath $\xi$}, \mbox{\boldmath $\xi$}_{\rm D})
         +\mbox{\boldmath $C$}^{\rm B}(\mbox{\boldmath $\xi$}, \mbox{\boldmath $\xi$}_{\rm D}).
\label{2.7'}
\end{equation}
A detailed expression for $\mbox{\boldmath $C$}^{\rm G}(\mbox{\boldmath $\xi$}, \mbox{\boldmath $\xi$}_{\rm D})$ is obtained by Kato
(2004, 2008) and duplicated in paper I.
One of major purposes of this paper is to derive $\mbox{\boldmath $C$}^{\rm B}(\mbox{\boldmath $\xi$}, \mbox{\boldmath $\xi$}_{\rm D})$.
To obtain $\mbox{\boldmath $C$}^{\rm B}$ in an explicit from of $\mbox{\boldmath $\xi$}$, we must express the Lagrangian variation of
magnetic force, $\Delta[(1/\rho){\rm curl} \mbox{\boldmath $B$}\times \mbox{\boldmath $B$}]$ until the second order terms
with respect to $\mbox{\boldmath $\xi$}$.
To do so, we need the Lagrangian variation of $\mbox{\boldmath $B$}$, i.e., $\Delta\mbox{\boldmath $B$}$, expressed until
the second order terms with respect to $\mbox{\boldmath $\xi$}$.
The procedures of these derivations and the results are, however, given in appendixes, since they are lengthy.
The derivation of $\Delta\mbox{\boldmath $B$}$ is in appendix 1, and that of 
$\Delta[(1/\rho){\rm curl} \mbox{\boldmath $B$}\times \mbox{\boldmath $B$}]$
is in appendix 2.

The disk oscillations, $\mbox{\boldmath $\xi$}(\mbox{\boldmath $r$}, t)$, resulting from the quasi-nonlinear coupling 
through disk deformation, $\mbox{\boldmath $\xi$}_{\rm D}(\mbox{\boldmath $r$}, t)$, will be written generally in the form (paper I)
\begin{eqnarray}
      \mbox{\boldmath $\xi$}(\mbox{\boldmath $r$},t)=&&\Re \sum_{{\rm i}=1}^2}A_{\rm i}(t)\hat{\mbox{\boldmath $\xi$}}_{\rm i}(\mbox{\boldmath $r$})
                       {\rm exp(i\omega_{\rm i} t)    
               +\Re \sum_i^{2}\sum_{\alpha\not= 1,2} A_{i,\alpha}\hat{\mbox{\boldmath $\xi$}}_\alpha(\mbox{\boldmath $r$}){\rm exp}(i\omega_{\rm i} t)
                             \nonumber   \\
             &&+ {\rm oscillating \ terms\ with \ other \ frequencies}.
\label{2.11}
\end{eqnarray}
The original two oscillations, $\mbox{\boldmath $\xi$}_1$ and $\mbox{\boldmath $\xi$}_2$, resonantly interact through the disk deformation.
Hence, their amplitudes secularly change with time, which is taken into account in equation (\ref{2.11}) by taking
the amplitudes, $A_{\rm i}$'s, to be slowly varying functions of time.
The terms whose time-dependence is exp($i\omega_{\rm i} t$) but whose spatial dependence is different from 
$\hat{\mbox{\boldmath $\xi$}}_{\rm i}$ are expressed by a sum of a series of eigen-functions, $\hat{\mbox{\boldmath $\xi$}}_\alpha$
 ($\alpha\not= 1$ and 2),
assuming that they make a complete set.
The terms  whose time-dependences are different from exp$(i\omega_{\rm i}t)$ are not written down explicitly in equation (\ref{2.11}),
since these terms disappear by taking long-term time average when we are interested in phenomena with frequencies 
of $\omega_{\rm i}$ (i $=$ 1 and 2).

In order to derive equations describing the time evolution of $A_{\rm i}(t)$, we substitute equation (\ref{2.11}) into the
left-hand side of equation (\ref{2.7}).
Then, considering that $\hat{\mbox{\boldmath $\xi$}}_{\rm i}$'s and $\hat{\mbox{\boldmath $\xi$}}_\alpha$'s are displacement vectors 
associated with eigen-functions of the linear wave equation (\ref{2.3}), we find that the left-hand side of equation (\ref{2.7}) 
becomes the real part of 
\begin{eqnarray}
   &&2\rho_0\sum_{i=1}^2\frac{dA_{\rm i}}{dt}\biggr[i\omega_{\rm i}+(\mbox{\boldmath $u$}_0\cdot\nabla)\biggr]
                     \hat{\mbox{\boldmath $\xi$}}_{\rm i}{\rm exp}(i\omega_{\rm i}t)
                    \nonumber       \\
   &&+\rho_0\sum_i\sum_{\alpha\not= 1,2} A_{i,\alpha}\biggr[(\omega_\alpha^2-\omega_{\rm i}^2)-2i(\omega_\alpha-\omega_{\rm i})(\mbox{\boldmath $u$}_0\cdot\nabla)\biggr]
         \hat{\mbox{\boldmath $\xi$}}_\alpha{\rm exp}(i\omega_{\rm i}t)    \nonumber  \\
   &&+ {\rm oscillating \ terms\ with \ other \ frequencies}.
\label{2.12}
\end{eqnarray}
where $d^2A_{\rm i}/d t^2$ has been neglected, since $A_{\rm i}(t)$($i=1$ and 2) are slowly varying functions of time.
Now, the real part of equation (\ref{2.12}) is integrated over the whole volume of disks after being multiplied by 
$\mbox{\boldmath $\xi$}_1[=\Re\ \hat{\mbox{\boldmath $\xi$}}_1{\rm exp}(i\omega_1t)]$.\footnote{
The formula
$$
      \Re(A)\Re(B)=\frac{1}{2}\Re[AB+AB^*]=\frac{1}{2}\Re[AB+A^*B]      \nonumber
$$
is used, where $A$ and $B$ are complex variables and $B^*$ is the complex conjugate of $B$.
}
Then, the term resulting from the second term of equation (\ref{2.12}) vanishes\footnote{See an orthogonal relation given by
equation (8) of Kato et al. (2011).
The relation holds even in the present case of hydromagnetic perturbations,
since what we use is only the fact that the operator $\mbox{\boldmath $L$}$ is Hermitian.
}
and the results become
\begin{equation}
      \Re \ i\frac{dA_1}{dt}\biggr\langle \rho_0\hat{\mbox{\boldmath $\xi$}}_1^*[\omega_1-i(\mbox{\boldmath $u$}_0\cdot\nabla)]
                \hat{\mbox{\boldmath $\xi$}}_1\biggr\rangle,
\label{2.13}
\end{equation}
which is further reduced to 
\begin{equation}
    \Re\  i\frac{2E_1}{\omega_1}\frac{dA_1}{dt},
\label{}
\end{equation}
where $E_1$ is the wave energy given by equation (\ref{2.14}).

After the above preparations, the real part of equation (\ref{2.7}) is multiplied by  
$\mbox{\boldmath $\xi$}_1(\mbox{\boldmath $r$})$ and integrated over the whole volume
to lead to 
\begin{equation}
     \Re \ i\frac{2E_1}{\omega_1}\frac{dA_1}{dt}
      =\frac{1}{2}\Re \ \biggr[A_2(t)A_{\rm D}\biggr\langle\hat{\mbox{\boldmath $\xi$}}_1\cdot
         \mbox{\boldmath $C$}(\hat{\mbox{\boldmath $\xi$}}_2,\hat{\mbox{\boldmath $\xi$}}_{\rm D})\biggr\rangle
            {\rm exp}(i\Delta \omega t)\biggr].
\label{2.15}
\end{equation}
In deriving the right-hand side of equation (\ref{2.15}), the time periodic terms with high frequencies (i.e., non-resonant terms) have
been neglected by time average being taken.

Similarly, we multiply $\mbox{\boldmath $\xi$}_2(\mbox{\boldmath $r$})$ to the real part of equation (\ref{2.7}) and integrate over the whole volume
to lead to
\begin{equation}
     \Re \ i\frac{2E_2}{\omega_1}\frac{dA_2}{dt}
      =\frac{1}{2}\Re \ \biggr[A_1(t)A_{\rm D}\biggr\langle\hat{\mbox{\boldmath $\xi$}}_2\cdot \mbox{\boldmath $C$}(\hat{\mbox{\boldmath $\xi$}}_1,
           \hat{\mbox{\boldmath $\xi$}}_{\rm D})\biggr\rangle
            {\rm exp}(i\Delta \omega t)\biggr].
\label{2.16}
\end{equation}
Equations (\ref{2.15}) and (\ref{2.16}) are formally the same as those in paper I,
but expressions for $\mbox{\boldmath $C$}$'s in the present case are generalizations of $\mbox{\boldmath $C$}$'s 
in paper I to hydromagnetic cases [see equation (\ref{2.7'})].

The most important characteristics of the coupling terms in equations (\ref{2.15}) and (\ref{2.16}) are that
$\hat{\mbox{\boldmath $\xi$}}_1$ and $\hat{\mbox{\boldmath $\xi$}}_2$ in
$\biggr\langle\hat{\mbox{\boldmath $\xi$}}_1\cdot
         \mbox{\boldmath $C$}(\hat{\mbox{\boldmath $\xi$}}_2,\hat{\mbox{\boldmath $\xi$}}_{\rm D})\biggr\rangle$ and 
$\biggr\langle\hat{\mbox{\boldmath $\xi$}}_2\cdot
         \mbox{\boldmath $C$}(\hat{\mbox{\boldmath $\xi$}}_1,\hat{\mbox{\boldmath $\xi$}}_{\rm D})\biggr\rangle$
are commutative.
That is 
\begin{equation}
      W\equiv\biggr\langle\hat{\mbox {\boldmath $\xi$}}_1\cdot\mbox {\boldmath $C$}
                         (\hat{\mbox {\boldmath $\xi$}}_2, \hat{\mbox {\boldmath $\xi$}}_{\rm D})\biggr\rangle
            =\biggr\langle\hat{\mbox {\boldmath $\xi$}}_2\cdot\mbox {\boldmath $C$}
                         (\hat{\mbox {\boldmath $\xi$}}_1, \hat{\mbox {\boldmath $\xi$}}_{\rm D})\biggr\rangle.
\label{3.3}
\end{equation}
A proof of this commutability is already given in Kato (2008) in the case of hydrodynamic oscillations.
Even in the case of hydromagnetic oscillations the commutability is generally realized.
We think there will be a simple way of the proof.\footnote{
We suppose the presence of a simple way of the proof, since the commutability may be a general characteristic of
conservative systems.}
However, we have not been able to construct a simple proof.
Hence, we must content with a clumsy analytical verification, which 
is very troublesome and thus given in appendix 3.
Basic assumptions involved in the proof is that when the volume integrations in equation (\ref{3.3}) are performed by parts,
the surface integrals vanish.

\section{Conditions of Resonant Growth and Discussions}

Since the equations describing time evolution of resonant oscillations, equations (\ref{2.15}) -- (\ref{2.16}), are
formally the same as those in paper I, we can derive the same conclusions in paper I.
When the amplitude $A_{\rm D}$ is assumed to be constant, 
what are governed by equations (\ref{2.15}) and (\ref{2.16}) are the imaginary part of $A_1$ and $A_2$, i.e., $A_{1,{\rm i}}$ 
and $A_{2,{\rm i}}$.
Their real parts are not related to the resonance, and we can neglect them in considering the exponential growth (or damping) of
$A_{1,{\rm i}}$ and $A_{2,{\rm i}}$. 
Then, restricting our attention only to the case of $\Delta\omega=0$,
we have from equations (\ref{2.15}) and (\ref{2.16})
\begin{equation}
     -\frac{2E_1}{\omega_1}\frac{dA_{1,{\rm i}}}{dt}=-\frac{1}{2}A_{2,{\rm i}}\Im (A_{\rm D}W),
\label{3.1}
\end{equation}
\begin{equation}
     -\frac{2E_2}{\omega_2}\frac{dA_{2,{\rm i}}}{dt}=-\frac{1}{2}A_{1,{\rm i}}\Im (A_{\rm D}W).
\label{3.2}
\end{equation}

Eliminating $A_{2,{\rm i}}$ from equations (\ref{3.1}) and (\ref{3.2}), we have
\begin{equation}
     \frac{d^2A_{1,{\rm i}}}{dt^2}
          =\frac{1}{16}\biggr(\frac{E_1E_2}{\omega_1\omega_2}\biggr)^{-1}\biggr[\Im (A_{\rm D}W)\biggr]^2A_{1,{\rm i}}.
\label{3.4}
\end{equation}
The same equation is obtained for $A_{2,{\rm i}}$ by eliminating $A_{1,{\rm i}}$ instead of $A_{2,{\rm i}}$.
Equation (\ref{3.4}) shows that if 
\begin{equation}
       \frac{E_1E_2}{\omega_1\omega_2}>0,
\label{3.5}
\end{equation}
the oscillations grow with the growth rate given by
\begin{equation}
      \biggr(\frac{\omega_1\omega_2}{16E_1E_2}\biggr)^{1/2}\vert \Im (A_{\rm D}W) \vert.
\label{3.6}
\end{equation}

The above results show that even in the case of resonant coupling of ideal MHD oscillations, the formal criterion of instability
obtained in paper I for hydrodynamic oscillations is unchanged.
That is, we have the following results.

1) Two oscillations with ($\omega_1$, $m_1$) and ($\omega_2$, $m_2$) in a deformed disk with ($\omega_{\rm D}$, $m_{\rm D}$)
are amplified if
\begin{equation}
       \frac{E_1}{\omega_1}\frac{E_2}{\omega_2}>0,
\label{3.7}
\end{equation}
where $\omega_1+\omega_2+\omega_{\rm D}=0$ and $m_1+m_2+m_{\rm D}=0$.

2) In the case where the frequency associated with the disk deformation, $\omega_{\rm D}$, is low, the resonant condition,
$\omega_1+\omega_2+\omega_{\rm D}=0$, is realized for $\omega_1\omega_2<0$.
Hence, in this case the amplification condition (\ref{3.7}) can be reduced to  
\begin{equation}
      E_1E_2<0.
\label{3.8}
\end{equation}
This might suggest that the resonant interaction between two oscillations with opposite signs of wave energy is
the cause of amplification.
However, in the case where $\omega_{\rm D}$ is so high that the resonant condition $\omega_1+\omega_2+\omega_{\rm D}=0$ is realized
for the same signs of $\omega_1$ and $\omega_2$, the amplification of the $\omega_1$- and $\omega_2$-oscillations occurs
when
\begin{equation}
       E_1E_2 >0.
\label{3.9}
\end{equation}
This is different from the case of $\omega_{\rm D}$ being small.
That is, the picture that the cause of resonant amplification is a direct energy exchange between two oscillations with 
positive and negative wave-energies is not always relevant.

3) The amplifications of $\omega_1$- and $\omega_2$-oscillations
will be generally interpreted as a result of energy exchange between the $\omega_{\rm D}$-oscillation and the set of
$\omega_1$- and $\omega_2$-oscillations.
This is based on the following consideration.
If the $\omega_{\rm D}$-oscillation is not maintained externally, unlike the treatment in this paper, 
equations (\ref{3.1}), (\ref{3.2}) and a similar equation for $A_{{\rm D, i}}$ [equation (38) in paper I] give
\begin{equation}
   \frac{E_1}{\omega_1}\frac{dA^2_{1,{\rm i}}}{dt}=\frac{E_2}{\omega_2}\frac{dA^2_{2,{\rm i}}}{dt}=
   \frac{E_{\rm D}}{\omega_{\rm D}}\frac{dA^2_{{\rm D},{\rm i}}}{dt}.
\label{3.10}
\end{equation}
Multiplying $E_1/\omega_1$ to these equations, we see that when the instability condition $(E_1/\omega_1)(E_2/\omega_2)>0$
is satisfied and thus $A_{1,{\rm i}}^2$ and $A_{2,{\rm i}}^2$ grow with time,  we have
\begin{equation}
      \frac{E_1}{\omega_1}\frac{E_{\rm D}}{\omega_{\rm D}}\frac{dA_{\rm D,i}^2}{dt}>0.
\label{3.11}
\end{equation}
This inequality means 
\begin{equation}
     \frac{dA_{\rm D,i}^2}{dt}<0,
\end{equation}
since $(E_1/\omega_1)(E_{\rm D}/\omega_{\rm D})$ is usually negative
in the case of $(E_1/\omega_1)(E_2/\omega_2)>0$ [see inequalities (41) and (42) in paper I].
The same conclusion can be derived by multiplying $E_2/\omega_2$ to equation (\ref{3.10}).
For more discussions, see equations (43) - (45) in paper I and the arguments following them in paper I.

In summary, wave energy, $E$, is an important concept in considering resonant amplification of oscillations,
but generally speaking, the more important quantity which is directly related to amplification is not $E$ but $E/\omega$.

It is important to notice here that simplicity of the instability condition, $(E_1/\omega_1)(E_2/\omega_2)>0$,  
comes from the commutability relation of the coupling term $W$.
The commutability is proofed in this paper by integrating the volume integrations in $W$ by
parts under neglect of the surface integrals (appendix 3).
This neglect of the surface integrals will be acceptable, if boundaries of the system are taken far outside 
and the displacement vectors associated with perturbations can be taken to be small there.
Such assumptions may not be always acceptable in realistic cases, and careful considerations may be necessary.   
We think, however, that the commutability relation is a general characteristic in conservative systems.
Hence, there will be a simple direct proof of the commutability,
which will give more physically perspective conditions of the commutability.
 
Deformed disks will appear in various astrophysical objects, e.g.,
in disks of binary systems and in disks where large scale instabilities are present.
In previous work, we have applied the present wave-wave 
resonant amplification process to describe two astrophysical phenomena.
The first one is the positive and negative superhumps in dwarf novae (Kato 2013a, b, c).\footnote{
See Osaki (1985) on the origin of the positive superhumps, and Lubow (1992) and references in Kato (2013c) on
possible origins of the negative superhumps.
}
This is one of examples where 
$\vert\omega_{\rm D}\vert$ is comparable with $\vert\omega_1\vert$ or $\vert\omega_2\vert$.
We are also trying, since Kato (2004), to apply the present wave-wave resonant process to  
the high frequency quasi-periodic oscillations (HFQPOs) observed in low-mass X-ray binaries (LMXBs).
This is an example of application of the wave-wave resonant process to a case where $\vert\omega_{\rm D}\vert$ is so small that 
$\omega_1+\omega_2+\omega_{\rm D}=0$ leads to $\omega_1\omega_2<0$ and the instability condition is reduced to
$E_1E_2<0$.
 
When the present wave-wave resonant process is applied to describe the oscillatory phenomena in disks, however,
magnetic fields are not always of importance.
In dwarf novae, for example, magnetic fields are not so strong that they will have no major effects on disk oscillations. 
In the case of LMXBs, however, this is not the case. 
In some microquasars, twin HFQPOs are observed with frequency ratio close to 3 : 2.
Furthermore, their central sources (black holes) have extremely high spins.
Any discoseismological models in which  effects of magnetic fields are not taken into account 
are not successful so far in describing simultaneously both the frequency ratio closed to 3 : 2 and 
the high spin of the central sources.
That is, magnetic fields will be important in describing the HFQPOs in microquasars by discoseismological models.

The importance of magnetic fields in considering discoseismology of black-hole sources partially comes from the fact
that the decrease of epicyclic frequency in the radial direction towards  the central source by
general relativistic effects (Okazaki et al. 1987) is strongly modified by the presence of poloidal magnetic fields, even if 
the fields do not have small plasma $\beta$-values (Fu \& Dai 2009).
Using this fact and assuming in advance that the instability condition $(E_1/\omega_1)(E_2/\omega_2)>0$
of wave-wave resonant process 
is unchanged even in the case of magnetized disks, Kato (2012) tried to describe the HFQPOs by the present
wave-wave resonant process, considering the resonant interaction between a set of a two-armed vertical p-mode oscillation 
and an axisymmetric g-mode oscillation in a disk with two-armed disk deformation. 
The results seem to be able to simultaneously describe both the frequency ratio and the high spin in conceivable ranges of 
strength of magnetic fields.
This suggests that further examinations of HFQPOs of black-hole sources along this line is worthwhile.

\bigskip\noindent
{\bf Appendix 1. Lagrangian and Eulerian Variations of Magnetic Fields until Second-Order Terms with Respect to 
Displacement Vector $\mbox{\boldmath $\xi$}$}

Our purpose here is to express the frozen-in condition of magnetic fields,
until the second-order terms with respect to displacement vector $\mbox{\boldmath $\xi$}$.
Let us consider a small rectangular solid in a fluid (see the left panel of figure 1).
Among eight vertices the cartesican coordinates of four ones, $A_0$, $B_0$, $C_0$, and $D_0$ are, respectively, 
$A_0(r_1,r_2,r_3)$, $B_0(r_1+{\tilde \Delta} r_1, r_2, r_3)$, $C_0(r_1,r_2+{\tilde \Delta} r_2, r_3)$, 
$D_0(r_1, r_2, r_3+{\tilde \Delta} r_3)$,
where ${\tilde \Delta} r_1$, ${\tilde \Delta} r_2$, ${\tilde \Delta} r_3$ are assumed to be sufficiently small.
In the unperturbed state the flux of magnetic fields passing through a piece of the rectangular plane specified by two
vectors $\stackrel{\longrightarrow}{A_0B_0}$ and $\stackrel{\longrightarrow}{A_0C_0}$ is $B_{03}{\tilde \Delta} r_1{\tilde \Delta} r_2$.
Similarly, the flux passing through the plane specified by $\stackrel{\longrightarrow}{A_0D_0}$ and
$\stackrel{\longrightarrow}{A_0B_0}$ is $B_{02}{\tilde \Delta} r_3{\tilde \Delta} r_1$, and the flux passing through the plane specified by
$\stackrel{\longrightarrow}{A_0C_0}$ and $\stackrel{\longrightarrow}{A_0D_0}$ is $B_{01}{\tilde \Delta} r_2{\tilde \Delta} r_3$.
Here, the first subspript 0 attached to magnetic field $B$ shows the unperturbed value and the second one does its spatial
component, and is not confused with the symbol of point $B_0$.

Now, we assume that a fluid element at point $A_0$ moves to point $A_1(a_1,a_2,a_3)$ (see the right panel of figure 1)
by a perturbation.
Here, $\mbox{\boldmath $a$}= (a_1, a_2, a_3)$ is related to $\mbox{\boldmath $r$}=(r_1,r_2,r_3)$ through the displacement vector
$\mbox{\boldmath $\xi$}=(\xi_1,\xi_2,\xi_3)$ as
\begin{equation}
       \mbox{\boldmath $a$}(\mbox{\boldmath $r$})=\mbox{\boldmath $r$}+\mbox{\boldmath $\xi$}(\mbox{\boldmath $r$}).
\label{deltaB.1}
\end{equation}
Corresponding to this displacement, a fluid element at point $B_0$ moves to point $B_1$, which is 
\begin{equation}
     B_1\biggr(a_1+\frac{\partial a_1}{\partial r_1}{\tilde \Delta} r_1, a_2+\frac{\partial a_2}{\partial r_1}{\tilde \Delta} r_1,
                   a_3+\frac{\partial a_3}{\partial r_1}{\tilde \Delta} r_1\biggr).
\label{deltaB.2}
\end{equation}
Similarly, fluid elements at $C_0$ and $D_0$ move, respectively, to points $C_1$ and $D_1$, which are
\begin{equation}
     C_1\biggr(a_1+\frac{\partial a_1}{\partial r_2}{\tilde \Delta} r_2, a_2+\frac{\partial a_2}{\partial r_2}{\tilde \Delta} r_2,
                   a_3+\frac{\partial a_3}{\partial r_2}{\tilde \Delta} r_2\biggr),
\label{deltaB.3}
\end{equation}
\begin{equation}
     D_1\biggr(a_1+\frac{\partial a_1}{\partial r_3}{\tilde \Delta} r_3, a_2+\frac{\partial a_2}{\partial r_3}{\tilde \Delta} r_3,
                   a_3+\frac{\partial a_3}{\partial r_3}{\tilde \Delta} r_3\biggr).
\label{deltaB.4}
\end{equation}
Hence, the vector defined by $\stackrel{\longrightarrow}{A_1B_1}(\equiv \tilde{\mbox{\boldmath $B$}})$,
$\stackrel{\longrightarrow}{A_1C_1}(\equiv \tilde{\mbox{\boldmath $C$}})$, and
$\stackrel{\longrightarrow}{A_1D_1}(\equiv \tilde{\mbox{\boldmath $D$}})$ are expressed, respectively, in the forms:    
\begin{equation}
   \tilde {\mbox{\boldmath $B$}}=\biggr(\frac{\partial a_1}{\partial r_1}, \frac{\partial a_2}{\partial r_1}, 
         \frac{\partial a_3}{\partial r_1}\biggr){\tilde \Delta} r_1,
\label{deltaB.5}
\end{equation}
\begin{equation}
   \tilde {\mbox{\boldmath $C$}}=\biggr(\frac{\partial a_1}{\partial r_2}, \frac{\partial a_2}{\partial r_2}, 
         \frac{\partial a_3}{\partial r_2}\biggr){\tilde \Delta} r_2,
\label{deltaB.6}
\end{equation}
\begin{equation}
   \tilde {\mbox{\boldmath $D$}}=\biggr(\frac{\partial a_1}{\partial r_3}, \frac{\partial a_2}{\partial r_3}, 
         \frac{\partial a_3}{\partial r_3}\biggr){\tilde \Delta} r_3.
\label{deltaB.7}
\end{equation}

\begin{figure}
\begin{center}
    \FigureFile(80mm,80mm){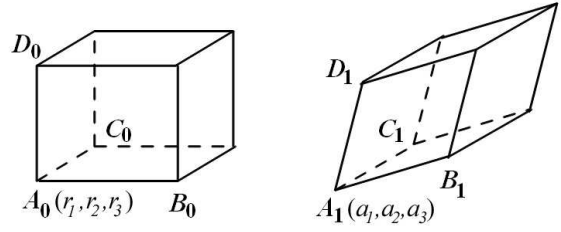}
\end{center}
\caption{A small rectangular solid element in fluid (left panel).
Its deformation after a perturbation (right panel).
}
\end{figure}

Then, $\tilde {\mbox{\boldmath $B$}}\times \tilde {\mbox{\boldmath $C$}}$ times $\mbox{\boldmath $B$}(\mbox{\boldmath $r$}
+\mbox{\boldmath $\xi$})$ [where $\mbox{\boldmath $B$}(\mbox{\boldmath $r$}+\mbox{\boldmath $\xi$})$ is 
the magnetic field at the displaced position, i.e., $\mbox{\boldmath $B$}(\mbox{\boldmath $r$}
+\mbox{\boldmath $\xi$})=\mbox{\boldmath $B$}_0(\mbox{\boldmath $r$})+\Delta\mbox{\boldmath $B$}$
and $\Delta \mbox{\boldmath $B$}$ is the Lagrangian variation\footnote{
Here and hereafter, the Lagrangian variation of $\mbox{\boldmath $B$}$ is denoted by $\Delta \mbox{\boldmath $B$}$,
and $\delta \mbox{\boldmath $B$}$ is reserved for the Eulerian variation of $\mbox{\boldmath $B$}$,
like Lynden-Bell and Ostriker (1967).
} 
of magnetic field $\mbox{\boldmath $B$}$] is
magnetic flux through the surface defined by vectors $\tilde{\mbox{\boldmath $B$}}$ and $\tilde{\mbox{\boldmath $C$}}$.
This flux must be equal to $B_{03}{\tilde \Delta} r_1{\tilde \Delta} r_2$, since the magnetic fields are assumed to be frozen in the fluid.
This frozen-in condition is written as
\begin{equation}
   \frac{1}{{\tilde \Delta} r_1{\tilde \Delta} r_2}(\tilde{\mbox{\boldmath $B$}}\times\tilde{\mbox{\boldmath $C$}})
             \cdot\mbox{\boldmath $B$}(\mbox{\boldmath $a$})
      =\left| \begin{array}{ccc}
          1+\partial\xi_1/\partial r_1 & \partial \xi_2/\partial r_1  & \partial \xi_3/\partial r_1 \\
          \partial\xi_1/\partial r_2   & 1+\partial \xi_2/\partial r_2 & \partial \xi_3/\partial r_2 \\
          B_{01}+\Delta B_1 & B_{02}+\Delta B_2 & B_{03}+\Delta B_3
   \end{array}
   \right|
   =B_{03}(\mbox{\boldmath $r$}).
\label{deltaB.8}
\end{equation}
Silimarly, we have
\begin{equation}
   \frac{1}{{\tilde \Delta} r_3{\tilde \Delta} r_1}(\tilde{\mbox{\boldmath $D$}}\times\tilde{\mbox{\boldmath $B$}})
                \cdot\mbox{\boldmath $B$}(\mbox{\boldmath $a$})
      =\left| \begin{array}{ccc}
          \partial\xi_1/\partial r_3 & \partial \xi_2/\partial r_3  & 1+\partial \xi_3/\partial r_3 \\
          1+\partial\xi_1/\partial r_1   & \partial \xi_2/\partial r_1 & \partial \xi_3/\partial r_1 \\
          B_{01}+\Delta B_1 & B_{02}+\Delta B_2 & B_{03}+\Delta B_3
   \end{array}
   \right|
   =B_{02}(\mbox{\boldmath $r$}),
\label{deltaB.9}
\end{equation}
\begin{equation}
   \frac{1}{{\tilde \Delta} r_2{\tilde \Delta} r_3}(\tilde{\mbox{\boldmath $C$}}\times\tilde{\mbox{\boldmath $D$}})
              \cdot\mbox{\boldmath $B$}(\mbox{\boldmath $a$})
      =\left| \begin{array}{ccc}
          \partial\xi_1/\partial r_2 & 1+\partial \xi_2/\partial r_2  & \partial \xi_3/\partial r_2 \\
          \partial\xi_1/\partial r_3   & \partial \xi_2/\partial r_3 & 1+\partial \xi_3/\partial r_3 \\
          B_{01}+\Delta B_1 & B_{02}+\Delta B_2 & B_{03}+\Delta B_3
   \end{array}
   \right|
   =B_{01}(\mbox{\boldmath $r$}).
\label{deltaB.10}
\end{equation}

Equations (\ref{deltaB.8}) -- (\ref{deltaB.10}) are a set of simultaneous algebraic equations with respect 
to $\Delta B_1$, $\Delta B_2$, and $\Delta B_3$.
Their expressions in terms of displacement vector $\mbox{\boldmath $\xi$}$ can be derived easily after
lengthy but straight-forward calculations, since what we need here are their successive forms expressed in terms of 
$\mbox{\boldmath $\xi$}$, i.e., their first- and second-order expressions with respect to $\mbox{\boldmath $\xi$}$.

After deriving the first-order expressions for $\Delta B_1$, $\Delta B_2$, and $\Delta B_3$, we can easily summarize
their expressions in a single vector form, which is\footnote{
Here and hereafter, we use Einstein's convention on indices, i.e.,
take summation if a term has the same index variable twice.
}
\begin{equation}
     (\Delta B_i)_1=B_{0j}\frac{\partial \xi_i}{\partial r_j}-B_{0i}{\rm div}\mbox{\boldmath $\xi$},
\label{deltaB.11}
\end{equation}
where $(\Delta B_i)_1$ denotes the first-order quantity of $\Delta B_i$.
Of course, we can easily derive this expression for $(\Delta B_i)_1$ from the induction equation by integrating 
it with respect to time.\footnote{
The induction equation, $\partial \mbox{\boldmath $B$}/\partial t={\rm curl}(\mbox{\boldmath $u$}\times
\mbox{\boldmath $B$})$, is rewitten as
$$
   \frac{d\mbox{\boldmath $B$}}{dt}={\rm curl}(\mbox{\boldmath $u$}\times \mbox{\boldmath $B$})+
       (\mbox{\boldmath $u$}\cdot\nabla)\mbox{\boldmath $B$}
       =(\mbox{\boldmath $B$}\cdot\nabla)\mbox{\boldmath $u$}-\mbox{\boldmath $B$}\ {\rm div}\mbox{\boldmath $u$}.
          \nonumber
$$
By integrating this equation with respect to time, we can obtain equation (\ref{deltaB.11}) 
as the first-order expression for $\Delta B_i$.
}
By substituting equation $(\Delta B_i)_1$ given above into equations 
(\ref{deltaB.8}) -- (\ref{deltaB.10}), we obtain the second-order expression for $\Delta B_i$,
i.e., $(\Delta B_i)_2$, which is 
\begin{equation}
     (\Delta B_i)_2=-B_{0j}\frac{\partial \xi_i}{\partial r_j}{\rm div}\mbox{\boldmath $\xi$}
         +\frac{1}{2}B_{0i}\biggr[\frac{\partial\xi_\ell}{\partial r_k}\frac{\partial \xi_k}{\partial r_\ell}
         +({\rm div}\mbox{\boldmath $\xi$})^2\biggr].
\label{deltaB.12}
\end{equation}

Next, let us summarize the second-order expression of the Eulerian variation of $\mbox{\boldmath $B$}$.
The Eulerian variation of $\mbox{\boldmath $B$}(\mbox{\boldmath $r$})$, i.e., $\delta \mbox{\boldmath $B$}$, is defined by
$\delta \mbox{\boldmath $B$}=\mbox{\boldmath $B$}(\mbox{\boldmath $r$})-\mbox{\boldmath $B$}_0(\mbox{\boldmath $r$})$.
In order to demonstrate explicitly the order with respect to $\mbox{\boldmath $\xi$}$, $\delta B_i(\mbox{\boldmath $r$})$
is written as  $(\delta B_i)_1 + (\delta B_i)_2 ...$,
where the subscripts 1 and 2 denote, respectively, the first- and second-order quantities with respect to $\mbox{\boldmath $\xi$}$,
i.e., 
\begin{equation}
    \delta B_i(\mbox{\boldmath $r$})\equiv B_i(\mbox{\boldmath $r$})-B_{0i}(\mbox{\boldmath $r$})
        =\biggr(\delta B_i(\mbox{\boldmath $r$})\biggr)_1+\biggr(\delta B_i(\mbox{\boldmath $r$})\biggr)_2+...
\label{commut.3}
\end{equation}

Since $\Delta B_i=B_i(\mbox{\boldmath $r$}+\mbox{\boldmath $\xi$}(\mbox{\boldmath $r$}))-B_{0i}(\mbox{\boldmath $r$})$,
we have, after taylor-expanding $B_i(\mbox{\boldmath $r$}+\mbox{\boldmath $\xi$})$ around $B_i(\mbox{\boldmath $r$})$, 
\begin{equation}
     (\Delta B_i)_1=(\delta B_i)_1+\xi_j\frac{\partial}{\partial r_j}B_{0i} 
\label{commut.4}
\end{equation}
in the first-order with respect to $\mbox{\boldmath $\xi$}$, and 
\begin{equation}
       (\Delta B_i)_2=(\delta B_i)_2+\xi_j\frac{\partial}{\partial r_j}(\delta B_i)_1+\frac{1}{2}\xi_\ell \xi_j\frac{\partial^2}
             {\partial r_\ell\partial r_j}B_{0i}
\label{commut.5}
\end{equation}
in the second-order with respect to $\mbox{\boldmath $\xi$}$.
By using expressions for $(\Delta B_i)_1$ and $(\Delta B_i)_2$ given, respectively, by equations (\ref{deltaB.11}) and (\ref{deltaB.12}), 
we have
\begin{equation}
       (\delta B_i)_1=\frac{\partial}{\partial r_j}\biggr[B_{0j}\xi_i-B_{0i}\xi_j\biggr],
\label{B_1}
\end{equation}
\begin{eqnarray}
      (\delta B_i)_2=&&-B_{0j}\frac{\partial\xi_i}{\partial r_j}{\rm div}\mbox{\boldmath $\xi$}+
              \frac{1}{2}B_{0i}\biggr[\frac{\partial\xi_\ell}{\partial r_k}\frac{\partial \xi_k}{\partial r_\ell}
                    +({\rm div}\mbox{\boldmath $\xi$})^2\biggr]-\xi_i\frac{\partial}{\partial r_j}(\delta B_i)_1   \nonumber  \\
            &&
               -\frac{1}{2}\xi_\ell\xi_j\frac{\partial^2}{\partial r_\ell\partial r_j}B_{0i}.
\label{B_2}
\end{eqnarray}

\bigskip\noindent
{\bf Appendix 2. A Quasi-Nonlinear Expression for Lagrangian Variation of Magnetic Force}

The hydromagnetic force per unit mass, $\mbox{\boldmath $F$}_{\rm B}$, is $\mbox{\boldmath $F$}_{\rm B}=(1/4\pi\rho){\rm curl}
\mbox{\boldmath $B$}\times\mbox{\boldmath $B$}$ and the $i$-component of its Lagrangian variation, $\Delta F_{{\rm B}i}$,
is written as
\begin{equation}
    \Delta F_{{\rm B}i}(\mbox{\boldmath $r$})=F_{{\rm B}i}(\mbox{\boldmath $r$}+\mbox{\boldmath $\xi$}(\mbox{\boldmath $r$}))
    -[F_{{\rm B}i}(\mbox{\boldmath $r$})]_0,
\label{commut.1}
\end{equation}
where $[X(\mbox{\boldmath $r$})]_0$ denotes the unperturbed value of $X(\mbox{\boldmath $r$})$, and is different
from $X(\mbox{\boldmath $r$})$, since the latter is the value of $X$ at $\mbox{\boldmath $r$}$ after perturbation.

First, we taylor-expand  $F_{{\rm B}i}(\mbox{\boldmath $r$}+\mbox{\boldmath $\xi$}(\mbox{\boldmath $r$}))$ around 
$F_{{\rm B}i}(\mbox{\boldmath $r$})$ until the second-order terms with respect to displacement vector $\mbox{\boldmath $\xi$}$.
Next, $F_{{\rm B}i}(\mbox{\boldmath $r$})$ is expanded as $[F_{{\rm B}i}(\mbox{\boldmath $r$})]_0
+[\delta F_{{\rm B}i}(\mbox{\boldmath $r$})]_1+[\delta F_{{\rm B}i}(\mbox{\boldmath $r$})]_2+....$.
By performing the first step we have
\begin{eqnarray}
     \Delta F_{{\rm B}i}=&&\frac{1}{4\pi\rho_0}\biggr(1+\xi_j\frac{\partial}{\partial x_j}+
     \frac{1}{2}\xi_j\xi_k\frac{\partial^2}{\partial r_j\partial r_k}\biggr)
     \biggr[-\frac{1}{2}\frac{\partial B^2}{\partial r_i}+\frac{\partial}{\partial r_\ell}(B_iB_\ell)\biggr]
           \nonumber \\
     &&-\frac{1}{4\pi\rho_0}\biggr(\frac{\Delta\rho}{\rho_0}\biggr)\biggr(1+\xi_j\frac{\partial}{\partial x_j}\biggr)
     \biggr[-\frac{1}{2}\frac{\partial B^2}{\partial r_i}+\frac{\partial}{\partial r_\ell}(B_iB_\ell)\biggr]
           \nonumber \\
     &&+\frac{1}{4\pi\rho_0}\biggr(\frac{\Delta\rho}{\rho_0}\biggr)^2
     \biggr[-\frac{1}{2}\frac{\partial B_0^2}{\partial r_i}+\frac{\partial}{\partial r_\ell}(B_{0i}B_{0\ell})\biggr]
           \nonumber \\
     &&-\frac{1}{4\pi\rho_0}
     \biggr[-\frac{1}{2}\frac{\partial B_0^2}{\partial r_i}+\frac{\partial}{\partial r_\ell}(B_{0i}B_{0\ell})\biggr],
\label{commut.2}
\end{eqnarray}
where $B^2=B_i(\mbox{\boldmath $r$})B_i(\mbox{\boldmath $r$})$.
To obtain an explicit form of $\Delta F_{{\rm B}i}$ untill the second order of $\mbox{\boldmath $\xi$}$, 
we must express the terms in the large brackets of the first and second lines of 
equation (\ref{commut.2}) until the first- or second-order terms with respect to $\mbox{\boldmath $\xi$}$.
To do so, we expand $B_i$ as $B_i=B_{0i}+(\delta B_i)_1+(\delta B_i)_2...$ and take into account that $(\delta B_i)_1$ and $(\delta B_i)_2$ are
given by equations (\ref{B_1}) and (\ref{B_2}). 
The linear part of $\Delta F_{{\rm B}i}$, i.e., $(\Delta F_{{\rm B}i})_1$,  is then given by
\begin{eqnarray}
     4\pi\rho_0(\Delta F_{{\rm B}i})_1
      =&&-\frac{\partial}{\partial r_i}\biggr[B_{0j}\frac{\partial}{\partial r_k}(B_{0k}\xi_j-B_{0j}\xi_k)\biggr] 
            \nonumber  \\
       &&+\frac{\partial}{\partial r_j}\biggr[B_{0i}\frac{\partial}{\partial r_k}(B_{0k}\xi_j-B_{0j}\xi_k)
                   +B_{0j}\frac{\partial}{\partial r_k}(B_{0k}\xi_i-B_{0i}\xi_k)\biggr]
                         \nonumber   \\
       &&+\biggr(\xi_k\frac{\partial}{\partial r_k}+{\rm div}\mbox{\boldmath $\xi$}\biggr) 
            \biggr[-\frac{1}{2}\frac{\partial B_0^2}{\partial r_i}+\frac{\partial}{\partial r_j}(B_{0i}B_{0j})\biggr].
\label{commut.6}
\end{eqnarray}

Writing down explicitly the second order terms of $\Delta F_{{\rm B}i}$ with respect to $\mbox{\boldmath $\xi$}$, 
i.e., $(\Delta F_{{\rm B}i})_2$, we have, from equation (\ref{commut.2}), 
\begin{eqnarray}
   4\pi\rho_0(\Delta F_{{\rm B}i})_2
       =&&-\frac{\partial}{\partial r_i}\biggr[B_{0j}(\delta B_j)_2+\frac{1}{2}(\delta B_j)^2_1 \biggr]
              \nonumber     \\
       && +\frac{\partial}{\partial r_j}\biggr[B_{0j}(\delta B_i)_2+B_{0i}(\delta B_j)_2+(\delta B_i)_1(\delta B_j)_1\biggr]
              \nonumber     \\
        &&+\xi_k\frac{\partial}{\partial r_k}\biggr[-\frac{\partial}{\partial r_i}\biggr( B_{0j}(\delta B_j)_1\biggr)+
                 \frac{\partial}{\partial r_j}\biggr(B_{0j}(\delta B_i)_1+B_{0i}(\delta B_j)_1\biggr)\biggr]
                     \nonumber      \\
        &&+\frac{1}{2}\xi_j\xi_k\frac{\partial^2}{\partial r_j\partial r_k}
               \biggr[-\frac{1}{2}\frac{\partial B_0^2}{\partial r_i}
                           +\frac{\partial}{\partial r_k}(B_{0k}B_{0i})\biggr]
                     \nonumber    \\
        &&+{\rm div} \mbox{\boldmath $\xi$}\biggr[-\frac{\partial}{\partial r_i}\biggr(B_{0j}(\delta B_j)_1\biggr)
             +\frac{\partial}{\partial r_j}\biggr(B_{0j}(\delta B_i)_1+B_{0i}(\delta B_j)_1\biggr)     \nonumber  \\
        && \hspace{60pt}              
             +\xi_j\frac{\partial}{\partial r_j}\biggr(-\frac{1}{2}\frac{\partial B_0^2}{\partial r_i}
                           +\frac{\partial}{\partial r_k}(B_{0k}B_{0i})\biggr)\biggr]
                     \nonumber     \\
        &&+\frac{1}{2}\biggr[({\rm div}\mbox{\boldmath $\xi$})^2-\frac{\partial \xi_\ell}{\partial r_k}
              \frac{\partial \xi_k}{\partial r_\ell}\biggr]
                  \biggr[-\frac{1}{2}\frac{\partial B_0^2}{\partial r_i}+\frac{\partial}{\partial r_k}(B_{0k}B_{0i})\biggr],
\label{commut.7}
\end{eqnarray} 
where 
\begin{equation}
      \frac{(\Delta\rho)_2}{\rho_0}=\frac{1}{2}\biggr[({\rm div}\mbox{\boldmath $\xi$})^2+
          \frac{\partial \xi_i}{\partial r_j}\frac{\partial \xi_j}{\partial r_i}\biggr]
\end{equation}
has been used (Kato 2004).
 
If Eulerian quantities, $(\delta B_i)_1$ and $(\delta B_i)_2$, in equation (\ref{commut.7}) are expressed in terms of 
$\mbox{\boldmath $\xi$}$ by using equations (\ref{B_1}) and (\ref{B_2}) we have an expression for $(\Delta F_{{\rm B}i})_2$ 
explicitly expressed in terms of $\mbox{\boldmath $\xi$}$.
The detailed expression for $(\Delta F_{{\rm B}i})_2$ is, however, unnecessary here.
What we need is commutative relations of the coupling terms resulting from $(\Delta F_{{\rm B}i})_2$, which are
discussed in the next appendix.

\bigskip\noindent
{\bf Appendix 3. Commutability of Magnetic Coupling Terms}

So far, we have attached subscripts 1, 2, and 3 (or $D$) to $\mbox{\boldmath $\xi$}$, like
$\mbox{\boldmath $\xi$}_1$, $\mbox{\boldmath $\xi$}_2$, and $\mbox{\boldmath $\xi$}_3$
(or $\mbox{\boldmath $\xi$}_{\rm D}$), in order to represent the set of three oscillation
modes.
In this appendix, however, the superscripts (1), (2), and (3) (or $({\rm D})$) are attached to
represent them, respectively, like $\mbox{\boldmath $\xi$}^{(1)}$, $\mbox{\boldmath $\xi$}^{(2)}$, and
$\mbox{\boldmath $\xi$}^{(3)}$ (or $\mbox{\boldmath $\xi$}^{({\rm D})})$, in order to
avoid confusions with other subscripts. 

Let us consider three modes of oscillations, $\mbox{\boldmath $\xi$}^{(1)}$, $\mbox{\boldmath $\xi$}^{(2)}$, and
$\mbox{\boldmath $\xi$}^{(3)}$, which resonantly couple with each other.
For simplicity, $\mbox{\boldmath $\xi$}^{(3)}$ is regarded as a mode of disk deformation, i.e.,
$\mbox{\boldmath $\xi$}^{({\rm D})}$,
and the oscillation modes of $\mbox{\boldmath $\xi$}^{(1)}$ and $\mbox{\boldmath $\xi$}^{(2)}$ resonantly couple 
with each other through $\mbox{\boldmath $\xi$}^{({\rm D})}$.

The wave equation describing $\mbox{\boldmath $\xi$}^{(1)}$ has quasi-nonlinear terms on the right-hand side of the equation
[cf., equation (\ref{2.7})], which consist of terms of two groups.
The first ones are coupling terms due to hydrodynamical processes, and the second ones are those
resulting from hydromagnetic couplings.
The former ones are already examined in paper I, and expressed there as $\mbox{\boldmath $C$}^{\rm T}
(\mbox{\boldmath $\xi$}_2, \mbox{\boldmath $\xi$}_{\rm D})$, including the case where the
deformation is maintained by external tidal force.
This term has been expressed as
$\mbox{\boldmath $C$}^{\rm G}(\mbox{\boldmath $\xi$}_2,\mbox{\boldmath $\xi$}_{\rm D})$ in 
section 3 of this paper, and
is $\mbox{\boldmath $C$}^{\rm G}(\mbox{\boldmath $\xi$}^{(2)},\mbox{\boldmath $\xi$}^{({\rm D})})$, if 
the notation of this appendix 3 is used.
The coupling terms of the latter group is 
$\rho_0 \Delta \mbox{\boldmath $F$}_{{\rm B}}(\mbox{\boldmath $\xi$}^{(2)}, \mbox{\boldmath $\xi$}^{({\rm D})})$, 
where the arguments of $\Delta \mbox{\boldmath $F$}_{{\rm B}}$ is written explicitly, since we are now interested in the coupling between
$\mbox{\boldmath $\xi$}^{(2)}$ and $\mbox{\boldmath $\xi$}^{({\rm D})}$ which feedbacks to $\mbox{\boldmath $\xi$}^{(1)}$.
Hereafter, we denote $\rho_0 \Delta \mbox{\boldmath $F$}_{{\rm B}}(\mbox{\boldmath $\xi$}^{(2)}, \mbox{\boldmath $\xi$}^{({\rm D})})$ by
$\mbox{\boldmath $C$}^{{\rm B}}(\mbox{\boldmath $\xi$}^{(2)}, \mbox{\boldmath $\xi$}^{({\rm D})})$, i.e.,
\begin{equation}
       \rho_0 \Delta \mbox{\boldmath $F$}_{{\rm B}}(\mbox{\boldmath $\xi$}^{(2)}, \mbox{\boldmath $\xi$}^{({\rm D})})
           =\mbox{\boldmath $C$}^{{\rm B}}(\mbox{\boldmath $\xi$}^{(2)}, \mbox{\boldmath $\xi$}^{({\rm D})}),
\label{defCB}
\end{equation}
which is nothing but 
$\mbox{\boldmath $C$}^{{\rm B}}(\mbox{\boldmath $\xi$}_{2}, \mbox{\boldmath $\xi$}_{\rm D})$
in section 3 of this paper.
The sum of $\mbox{\boldmath $C$}^{{\rm G}}(\mbox{\boldmath $\xi$}^{(2)}, \mbox{\boldmath $\xi$}^{({\rm D})})$
and $\mbox{\boldmath $C$}^{{\rm B}}(\mbox{\boldmath $\xi$}^{(2)}, \mbox{\boldmath $\xi$}^{({\rm D})})$ is the
total coupling terms [see equation (\ref{2.7'}) in section 3], i.e., in the present notation we have
\begin{equation}
    \mbox{\boldmath $C$}(\mbox{\boldmath $\xi$}^{(2)}, \mbox{\boldmath $\xi$}^{({\rm D})})
    =\mbox{\boldmath $C$}^{({\rm G})}(\mbox{\boldmath $\xi$}^{(2)}, \mbox{\boldmath $\xi$}^{({\rm D})})
    +\mbox{\boldmath $C$}^{({\rm B})}(\mbox{\boldmath $\xi$}^{(2)}, \mbox{\boldmath $\xi$}^{({\rm D})}).
\label{add}
\end{equation}
It is noted that $\mbox{\boldmath $\xi$}^{(2)}$ and $\mbox{\boldmath $\xi$}^{({\rm D})}$ are commutative
in $\mbox{\boldmath $C$}^{({\rm B})}$, i.e., 
$\mbox{\boldmath $C$}^{{\rm B}}(\mbox{\boldmath $\xi$}^{(2)}, \mbox{\boldmath $\xi$}^{({\rm D})})=
\mbox{\boldmath $C$}^{{\rm B}}(\mbox{\boldmath $\xi$}^{({\rm D})}, \mbox{\boldmath $\xi$}^{(2)})$.

The purpose of this appendix is to show that in the volume integration of
\begin{equation}
      \int \mbox{\boldmath $\xi$}^{(1)}\cdot\mbox{\boldmath $C$}(\mbox{\boldmath $\xi$}^{(2)}, 
           \mbox{\boldmath $\xi$}^{({\rm D})})dV,
\label{C.1}
\end{equation}
$\mbox{\boldmath $\xi$}^{(1)}$ and $\mbox{\boldmath $\xi$}^{(2)}$ are exchangeable as
\begin{equation}
   \int\mbox{\boldmath $\xi$}^{(1)}\cdot\mbox{\boldmath $C$}(\mbox{\boldmath $\xi$}^{(2)}, \mbox{\boldmath $\xi$}^{({\rm D})})dV
  =\int\mbox{\boldmath $\xi$}^{(2)}\cdot\mbox{\boldmath $C$}(\mbox{\boldmath $\xi$}^{(1)}, \mbox{\boldmath $\xi$}^{({\rm D})})dV,
\label{C.2}
\end{equation}
where the integration is performed over the whole volume of the system.
A basic assumption involved in this derivation of commutability is that the surface integrals which appear by performing the integration
by part can be neglected.
The commutability of the hydrodynamical terms has been shown in paper I.
Hence, in this appendix we show only the commutability of hydromagnetic coupling terms: i.e., 
\begin{equation}
   \int\mbox{\boldmath $\xi$}^{(1)}\cdot\mbox{\boldmath $C$}^{({\rm B})}(\mbox{\boldmath $\xi$}^{(2)}, \mbox{\boldmath $\xi$}^{({\rm D})})dV
  =\int\mbox{\boldmath $\xi$}^{(2)}\cdot\mbox{\boldmath $C$}^{({\rm B})}(\mbox{\boldmath $\xi$}^{(1)}, \mbox{\boldmath $\xi$}^{({\rm D})})dV.
\label{C.2'}
\end{equation}

Let us multiply $\mbox{\boldmath $\xi$}^{(1)}$ to equation (\ref{commut.7}) and integrate by part, assuming that the surface integrals 
can be neglected.
Then, some terms resulting from the third and fifth lines of equations (\ref{commut.7}) are cancelled out, and we have
\begin{equation}
    4\pi\int \mbox{\boldmath $\xi$}^{(1)}\cdot\mbox{\boldmath $C$}^{\rm B}(\mbox{\boldmath $\xi$}^{(2)}, \mbox{\boldmath $\xi$}^{({\rm D})})dV
         =\int G^{\rm B}(\mbox{\boldmath $\xi$}^{(1)}, \mbox{\boldmath $\xi$}^{(2)}, \mbox{\boldmath $\xi$}^{({\rm D})})dV,
\label{C.6}
\end{equation}
where
\begin{eqnarray}
     G^{\rm B}
     =&&{\rm div}\mbox{\boldmath $\xi$}^{(1)}\biggr[B_{0j}(\delta B_j)_2+\frac{1}{2}(\delta B_j)_1^2\biggr] 
                   \nonumber \\
               &&-\frac{\partial \xi_i^{(1)}}{\partial r_j}\biggr[B_{0i}(\delta B_j)_2+B_{0j}(\delta B_i)_2+(\delta B_i)_1(\delta B_j)_1\biggr]
                   \nonumber  \\    
               &&+\frac{\partial \xi_i^{(1)}}{\partial r_j}\xi_j\biggr[\frac{\partial}{\partial r_i}\biggr(B_{0k}(\delta B_k)_1\biggr)
                    -\frac{\partial}{\partial r_k}\biggr(B_{0k}(\delta B_i)_1+B_{0i}(\delta B_k)_1\biggr) \biggr]
                   \nonumber \\
               &&+\frac{1}{2}\xi_i^{(1)}\xi_j\xi_k\frac{\partial^2}{\partial r_j\partial r_k}
                    \biggr[-\frac{1}{2}\frac{\partial B_0^2}{\partial r_i}
                    +\frac{\partial}{\partial r_\ell}(B_{0i}B_{0\ell})\biggr]
                        \nonumber     \\
               &&+\xi_i^{(1)}\xi_j{\rm div}\mbox{\boldmath $\xi$}\frac{\partial}{\partial r_j}
                     \biggr[-\frac{1}{2}\frac{\partial B_0^2}{\partial r_i}
                     +\frac{\partial}{\partial r_\ell}(B_{0i}B_{0\ell})\biggr]
                          \nonumber    \\
               &&+\frac{1}{2}\xi_i^{(1)}\biggr[({\rm div}\mbox{\boldmath $\xi$})^2 -\frac{\partial\xi_\ell}{\partial r_k}
                     \frac{\partial \xi_k}{\partial r_\ell}\biggr]
                     \biggr[-\frac{1}{2}\frac{\partial B_0^2}{\partial r_i}
                     +\frac{\partial}{\partial r_j}(B_{0i}B_{0j})\biggr].
\label{C.7}
\end{eqnarray}
In the above expression for $G^{\rm B}$, $\mbox{\boldmath $\xi$}^{(1)}$ is written explicitly, 
but $\mbox{\boldmath $\xi$}^{(2)}$ and $\mbox{\boldmath $\xi$}^{({\rm D})}$ are not.
Each term of equation (\ref{C.7}) has such a form as a component of $\mbox{\boldmath $\xi$}^{(1)}$ (or its spatial derivative)
times a quadratic term consisting of a component of $\mbox{\boldmath $\xi$}^{(2)}$ (or its spatial derivative) and 
a component of $\mbox{\boldmath $\xi$}^{({\rm D})}$ (or its derivative), 
although some terms are not explicitly expressed in such forms.
Each quadratic term mentioned above is the sum of two terms where the part of $\mbox{\boldmath $\xi$}^{(2)}$ and  
$\mbox{\boldmath $\xi$}^{({\rm D})}$ are changed.
With this convention we have neglected attaching the superscript $(2)$ or $({\rm D})$ in terms related to  
$\mbox{\boldmath $\xi$}$ in equation (\ref{C.7}),
in order to avoid complexity of notations.   
This convention is used hereafter.

The terms $(\delta B_j)_2$ and $(\delta B_i)_2$ in equation (\ref{C.7}) (there are three terms in all) can be expressed as
$(\delta B_j)_2=-\xi_\ell \partial (\delta B_j)_1/\partial r_\ell+...$ and  $(\delta B_i)_2=-\xi_\ell\partial (\delta B_i)_1/\partial r_\ell+...$ as shown in 
equation (\ref{B_2}).
These terms with $-\xi_\ell\partial (\delta B_j)_1/\partial r_\ell$ or $-\xi_\ell\partial (\delta B_i)_1/\partial r_\ell$ 
in equation (\ref{C.7}) are integrated by part in equation (\ref{C.6}), neglecting the surface integrals.
Then, contribution of the terms in $G^{\rm B}$ become as
\begin{equation}
     (\delta B_j)_1\frac{\partial}{\partial r_k}\biggr(B_{0j}\xi_k{\rm div}\mbox{\boldmath $\xi$}^{(1)}
              -B_{0i}\xi_k\frac{\partial\xi_i^{(1)}}{\partial r_j}\biggr)
       -(\delta B_i)_1\frac{\partial}{\partial r_k}\biggr(B_{0j}\xi_k\frac{\partial\xi_i^{(1)}}{\partial r_j}\biggr).
\label{C.7'}
\end{equation}
Furthermore, all the terms of the third line of equation (\ref{C.7}) are integrated by part in order to erase 
the terms with spatial derivatives of $(\delta B_j)_1$ and $(\delta B_i)_1$.
The result is
\begin{eqnarray}
    &&   -\biggr(\xi_j\frac{\partial}{\partial r_j}{\rm div}\mbox{\boldmath $\xi$}^{(1)}
          +\frac{\partial \xi_i^{(1)}}{\partial r_j}\frac{\partial \xi_j}{\partial r_i}\biggr)\biggr(B_{0k}(\delta B_k)_1\biggr)
                   \nonumber   \\
    && +\biggr(\frac{\partial^2\xi_i^{(1)}}{\partial r_j\partial r_k}\xi_j
          +\frac{\partial \xi_i^{(1)}}{\partial r_j}\frac{\partial\xi_j}{\partial r_k}\biggr)
                \biggr(B_{0k}(\delta B_i)_1+B_{0i}(\delta B_k)_1\biggr).
\label{C.8}
\end{eqnarray}
The sum of equations (\ref{C.7'}) and (\ref{C.8}) becomes, after some manipulations,
\begin{eqnarray}
      &&{\rm div}\mbox{\boldmath $\xi$}^{(1)}(\delta B_j)_1\frac{\partial}{\partial r_k}(\xi_kB_{0j})    \nonumber    \\
      &&+\frac{\partial \xi_i^{(1)}}{\partial r_j}\biggr[-(\delta B_j)_1\frac{\partial}{\partial r_k}(\xi_kB_{0i})+(\delta B_i)_1(\delta B_j)_1
           +(\delta B_k)_1\biggr(-B_{0k}\frac{\partial\xi_j}{\partial r_i}+B_{0i}\frac{\partial\xi_j}{\partial r_k}\biggr)\biggr].
\label{C.8'}
\end{eqnarray}
The remaining terms in the first and second lines of equation (\ref{C.7}) are arranged as, using equation (\ref{commut.5}), 
\begin{eqnarray}
     &&{\rm div}\mbox{\boldmath $\xi$}^{(1)}\biggr[B_{0j}\biggr((\Delta B_j)_2
      -\frac{1}{2}\xi_\ell\xi_k\frac{\partial^2 B_{0j}}{\partial r_\ell\partial r_k}\biggr)+\frac{1}{2}(\delta B_j)_1^2\biggr]
            \nonumber    \\
     &&-\frac{\partial \xi_i^{(1)}}{\partial r_j}\biggr[B_{0i}(\Delta B_j)_2+B_{0j}(\Delta B_i)_2+(\delta B_i)_1(\delta B_j)_1  \nonumber  \\         
     &&\hspace{35pt} -\frac{1}{2}\xi_\ell\xi_k\biggr(B_{0i}\frac{\partial^2 B_{0j}}{\partial r_\ell\partial r_k}
                +B_{0j}\frac{\partial^2 B_{0i}}{\partial r_\ell\partial r_k}\biggr)\biggr].
\label{C.8''}
\end{eqnarray}
Summing equations (\ref{C.8'}), (\ref{C.8''}), and the remaining terms in equation (\ref{C.7}), 
we finally obtain, after some manipulations,  an expression for $G^{\rm B}$
expressed explicitly in quadratic forms with respect to $\mbox{\boldmath $\xi$}$:     
\begin{eqnarray}
    G^{\rm B}\Rightarrow &&{\rm div} \mbox{\boldmath $\xi$}^{(1)}\Biggr[-B_{0j}B_{0\ell}\frac{\partial\xi_j}{\partial r_\ell}{\rm div}\mbox{\boldmath $\xi$}
                   +\frac{1}{2}B_0^2\biggr(\frac{\partial \xi_\ell}{\partial r_k}\frac{\partial\xi_k}{\partial r_\ell}
                   +({\rm div} \mbox{\boldmath $\xi$})^2\biggr)
                          \nonumber   \\
              &&\hspace{30pt}+\frac{1}{2}\frac{\partial\xi_j}{\partial r_\ell}\frac{\partial\xi_j}{\partial r_k}B_{0\ell}B_{0k}
                -\frac{1}{2}\biggr(B_{0j}{\rm div}\mbox{\boldmath $\xi$}+\xi_\ell\frac{\partial B_{0j}}{\partial r_\ell}\biggr)^2
                -\frac{1}{2}\xi_k\xi_\ell B_{0j}\frac{\partial^2 B_{0j}}{\partial r_k\partial r_\ell} \Biggr]    
                          \nonumber     \\
    &&\hspace{-7pt}+\frac{\partial\xi_i^{(1)}}{\partial r_j}\Biggr[\frac{\partial\xi_i}{\partial r_\ell}{\rm div}\mbox{\boldmath $\xi$}\ B_{0j}B_{0\ell}
                -\frac{\partial\xi_\ell}{\partial r_k}\frac{\partial\xi_k}{\partial r_\ell}B_{0i}B_{0j}
                           \nonumber      \\
               &&\hspace{30pt}+\xi_k{\rm div}\mbox{\boldmath $\xi$}\frac{\partial}{\partial r_k}\biggr(B_{0i}B_{0j}\biggr)
                 +\frac{1}{2}\xi_k\xi_\ell\frac{\partial^2}{\partial r_\ell\partial r_k}\biggr(B_{0i}B_{0j}\biggr)
                           \nonumber    \\
               &&\hspace{30pt}
                     -\frac{\partial\xi_j}{\partial r_i}B_{0k}\frac{\partial}{\partial r_\ell}
                        \biggr(B_{0\ell}\xi_k-B_{0k}\xi_\ell\biggr)\Biggr]
                             \nonumber      \\
               &&+\frac{\partial\xi_i^{(1)}}{\partial r_j}\frac{\partial\xi_j}{\partial r_k}
                          \biggr[-\xi_\ell B_{0k}\frac{\partial B_{0i}}{\partial r_\ell}
                         +B_{0i}\frac{\partial}{\partial r_\ell}(B_{0\ell}\xi_k-B_{0k}\xi_\ell)\biggr]
                             \nonumber    \\
               &&+\frac{1}{2}\xi_i^{(1)}\xi_j\xi_k\frac{\partial^2}{\partial r_j\partial r_k}
                     \biggr[-\frac{1}{2}\frac{\partial B_0^2}{\partial r_i}+\frac{\partial}{\partial r_\ell}(B_{0i}B_{0\ell})\biggr]
                                \nonumber           \\
               &&+\xi_i^{(1)}\xi_j{\rm div}\mbox{\boldmath $\xi$}\frac{\partial}{\partial r_j}
                     \biggr[-\frac{1}{2}\frac{\partial B_0^2}{\partial r_i}
                     +\frac{\partial}{\partial r_\ell}(B_{0i}B_{0\ell})\biggr]
                          \nonumber    \\
               &&+\frac{1}{2}\xi_i^{(1)}\biggr[({\rm div}\mbox{\boldmath $\xi$})^2 -\frac{\partial\xi_\ell}{\partial r_k}
                     \frac{\partial \xi_k}{\partial r_\ell}\biggr]
                     \biggr[-\frac{1}{2}\frac{\partial B_0^2}{\partial r_i}
                     +\frac{\partial}{\partial r_j}(B_{0i}B_{0j})\biggr].
\label{C.10}
\end{eqnarray}

Now, we introduce the following scalar quantities:
\begin{equation}
     {\cal A}(\mbox{\boldmath $\xi$}^{(1)}, \mbox{\boldmath $\xi$}^{(2)}, \mbox{\boldmath $\xi$}^{({\rm D})})
          \equiv B_0^2\biggr[\frac{1}{2}{\rm div}\mbox{\boldmath $\xi$}^{(1)}
                  \frac{\partial\xi_i}{\partial r_j}\frac{\partial\xi_j}{\partial r_i}
                  +{\rm div}\mbox{\boldmath $\xi$}\frac{\partial\xi_i^{(1)}}{\partial r_j}\frac{\partial\xi_j}{\partial r_i}\biggr],
\label{C.11}
\end{equation}
\begin{equation}
     {\cal B}_1(\mbox{\boldmath $\xi$}^{(1)}, \mbox{\boldmath $\xi$}^{(2)}, \mbox{\boldmath $\xi$}^{({\rm D})})
         \equiv \frac{1}{2}\frac{\partial B_0^2}{\partial r_k}\biggr[\frac{1}{2}\xi_k^{(1)}\frac{\partial \xi_i}{\partial r_j}
           \frac{\partial\xi_j}{\partial r_i}
           +\xi_k\frac{\partial\xi_i^{(1)}}{\partial r_j}\frac{\partial\xi_j}{\partial r_i}\biggr],
\label{C.12}
\end{equation}
\begin{equation}
     {\cal B}_2(\mbox{\boldmath $\xi$}^{(1)}, \mbox{\boldmath $\xi$}^{(2)}, \mbox{\boldmath $\xi$}^{({\rm D})})
         \equiv -\frac{1}{2}\frac{\partial B_0^2}{\partial r_k}\biggr[\frac{1}{2}\xi_k^{(1)}({\rm div}\mbox{\boldmath $\xi$})^2
             +\xi_k{\rm div}\mbox{\boldmath $\xi$}^{(1)}{\rm div}\mbox{\boldmath $\xi$}\biggr],
\label{C.13}
\end{equation}
\begin{equation}
      {\cal C}(\mbox{\boldmath $\xi$}^{(1)}, \mbox{\boldmath $\xi$}^{(2)}, \mbox{\boldmath $\xi$}^{({\rm D})})
          \equiv -\frac{1}{2}\frac{\partial^2B_0^2}{\partial r_i\partial r_j}\biggr[\frac{1}{2}\xi_i\xi_j{\rm div}\mbox{\boldmath $\xi$}^{(1)}
              +\xi_i^{(1)}\xi_j{\rm div}\mbox{\boldmath $\xi$}\biggr],
\label{C.14}
\end{equation}
\begin{equation}
      {\cal D}(\mbox{\boldmath $\xi$}^{(1)}, \mbox{\boldmath $\xi$}^{(2)}, \mbox{\boldmath $\xi$}^{({\rm D})})
           \equiv -\frac{1}{4}\frac{\partial^2B_0^2}{\partial r_i\partial r_j\partial r_k}\xi_i^{(1)}\xi_j\xi_k.
\label{C.14add}
\end{equation}
The right-hand sides of the above equations are written by use of the convention mentioned just below equation (\ref{C.7}).
That is, each term on the right-hand sides is quadratic in such a sense that one of a component of 
$\mbox{\boldmath $\xi$}$ (or its derivative) is that of $\mbox{\boldmath $\xi$}^{(2)}$ (or its detivative) and
the other is a component of $\mbox{\boldmath $\xi$}^{({\rm D})}$ (or its derivative).
The inverse case is also involved.
That is, the term is involved where the former component of $\mbox{\boldmath $\xi$}$ is 
that of $\mbox{\boldmath $\xi$}^{({\rm D})}$
(or its derivative) and the latter is that of $\mbox{\boldmath $\xi$}^{(2)}$ (or its derivative).
An important characteristic of the quantities defined by equations (\ref{C.11}) -- (\ref{C.14}) are that they are 
commutative to exchange of $\mbox{\boldmath $\xi$}^{(1)}$, $\mbox{\boldmath $\xi$}^{(2)}$, and $\mbox{\boldmath $\xi$}^{(3)}$.
For example, we have
\begin{equation}
        {\cal A}(\mbox{\boldmath $\xi$}^{(1)}, \mbox{\boldmath $\xi$}^{(2)}, \mbox{\boldmath $\xi$}^{({\rm D})})
       = {\cal A}(\mbox{\boldmath $\xi$}^{(2)}, \mbox{\boldmath $\xi$}^{(1)}, \mbox{\boldmath $\xi$}^{({\rm D})})
       = {\cal A}(\mbox{\boldmath $\xi$}^{({\rm D})}, \mbox{\boldmath $\xi$}^{(2)}, \mbox{\boldmath $\xi$}^{(1)})
       = ...
\label{C.14'}
\end{equation}
        
Then, the terms which are proportional to $B_0^2$ or to its spatial derivatives on the right-hand side of equation (\ref{C.10}) 
are all summarized by using ${\cal A}_1$, ${\cal B}_1$, ${\cal B}_2$, 
${\cal C}$, and ${\cal D}$.
$G^{\rm B}$ is then reduced to
\begin{eqnarray}
    G^{\rm B}\Rightarrow&& {\cal A}+{\cal B}_1+{\cal B}_2+{\cal C}+{\cal D}   \nonumber \\
              &&+{\rm div} \mbox{\boldmath $\xi$}^{(1)}\Biggr[-B_{0j}B_{0\ell}\frac{\partial\xi_j}{\partial r_\ell}{\rm div}\mbox{\boldmath $\xi$}
              +\frac{1}{2}B_{0\ell}B_{0k}\frac{\partial\xi_j}{\partial r_\ell}\frac{\partial\xi_j}{\partial r_k}\Biggr]    
                          \nonumber     \\
    &&\hspace{-7pt}+\frac{\partial\xi_i^{(1)}}{\partial r_j}\Biggr[B_{0j}B_{0\ell}\frac{\partial\xi_i}{\partial r_\ell}{\rm div}\mbox{\boldmath $\xi$}
                -B_{0i}B_{0j}\frac{\partial\xi_\ell}{\partial r_k}\frac{\partial\xi_k}{\partial r_\ell}
                -B_{0k}B_{0\ell}\frac{\partial\xi_j}{\partial r_i}\frac{\partial \xi_k}{\partial r_\ell}
                           \nonumber      \\
               &&\hspace{30pt}+\xi_k{\rm div}\mbox{\boldmath $\xi$}\frac{\partial}{\partial r_k}(B_{0i}B_{0j})
                 +\frac{1}{2}\xi_k\xi_\ell\frac{\partial^2}{\partial r_\ell\partial r_k}(B_{0i}B_{0j})\Biggr]
                           \nonumber    \\
    &&+\frac{\partial\xi_i^{(1)}}{\partial r_j}\frac{\partial\xi_j}{\partial r_k}
                          \biggr[-\xi_\ell B_{0k}\frac{\partial B_{0i}}{\partial r_\ell}
                         +B_{0i}\frac{\partial}{\partial r_\ell}(B_{0\ell}\xi_k-B_{0k}\xi_\ell)\biggr]
                             \nonumber    \\
    &&+\frac{1}{2}\xi_i^{(1)}\xi_j\xi_k\frac{\partial^3}{\partial r_j\partial r_k\partial r_\ell}(B_{0i}B_{0\ell})
                +\xi_i^{(1)}\xi_j{\rm div}\mbox{\boldmath $\xi$}\frac{\partial^2}{\partial r_j\partial r_\ell}(B_{0i}B_{0\ell})
                          \nonumber    \\
    &&+\frac{1}{2}\xi_i^{(1)}\biggr[({\rm div}\mbox{\boldmath $\xi$})^2 -\frac{\partial\xi_\ell}{\partial r_k}
                     \frac{\partial \xi_k}{\partial r_\ell}\biggr]
                     \frac{\partial}{\partial r_j}(B_{0i}B_{0j}).
\label{C.10_1}
\end{eqnarray}

Next, let us consider the terms proportional to $B_{0i}B_{0j}$ on the right-hand side of equation (\ref{C.10_1}).
The sum of the second term of the second line of equation (\ref{C.10_1}) and the first term of the third line is summarized as 
${\cal A}_{{\rm c}1}$:
\begin{equation}
     {\cal A}_{{\rm c}1}(\mbox{\boldmath $\xi$}^{(1)}, \mbox{\boldmath $\xi$}^{(2)}, \mbox{\boldmath $\xi$}^{({\rm D})})
         \equiv B_{0i}B_{0j}\biggr[\frac{1}{2}{\rm div}\mbox{\boldmath $\xi$}^{(1)}\frac{\partial \xi_k}{\partial r_i}
          \frac{\partial \xi_k}{\partial r_j}
          +\frac{\partial\xi_k^{(1)}}{\partial r_i}{\rm div}\mbox{\boldmath $\xi$}\frac{\partial\xi_k}{\partial r_j}\biggr],
\label{C.15}
\end{equation}
where the right-hand side of equation (\ref{C.15}) is written by use of the convention mentioned before 
[see sentences below equation (\ref{C.7}).
(This conventional expression is used hereafter for some scalar quantities without notice.) 
It is important to note that ${\cal A}_{{\rm c}1}(\mbox{\boldmath $\xi$}^{(1)}, \mbox{\boldmath $\xi$}^{(2)}, \mbox{\boldmath $\xi$}^{({\rm D})})$
is commutative with respect to $\mbox{\boldmath $\xi$}^{(1)}$, $\mbox{\boldmath $\xi$}^{(2)}$, and $\mbox{\boldmath $\xi$}^{({\rm D})}$.
In equation (\ref{C.10_1}) the remaining terms which have $B_{0i}B_{0j}$ with no spatial derivatives are
\begin{eqnarray}
      &&B_{0i}B_{0j}\biggr[-{\rm div}\mbox{\boldmath $\xi$}^{(1)}\frac{\partial\xi_i}{\partial r_j}{\rm div}\mbox{\boldmath $\xi$}
         -\frac{\partial\xi_i^{(1)}}{\partial r_j}\frac{\partial\xi_\ell}{\partial r_k}\frac{\partial\xi_k}{\partial r_\ell}
         -\frac{\partial\xi_k^{(1)}}{\partial r_\ell}\frac{\partial\xi_i}{\partial r_j}\frac{\partial\xi_\ell}{\partial r_k}
                \nonumber   \\
       &&\hspace{40pt} +\frac{\partial\xi_i^{(1)}}{\partial r_\ell}\frac{\partial \xi_\ell}{\partial r_k}\frac{\partial \xi_k}{\partial r_j}
          -\frac{\partial\xi_i^{(1)}}{\partial r_k}\frac{\partial\xi_k}{\partial r_j}{\rm div}\mbox{\boldmath $\xi$}\biggr].
\label{C.16}
\end{eqnarray}
The final term of equation (\ref{C.16}) can be reformed by integrating by parts, neglecting the surface integral, as
\begin{equation}
    B_{0i}B_{0j}\biggr[\frac{1}{2}\xi_i^{(1)}\frac{\partial}{\partial r_j}({\rm div}\mbox{\boldmath $\xi$})^2
         +\xi_i^{(1)}\frac{\partial\xi_k}{\partial r_j}\frac{\partial}{\partial r_k}{\rm div}\mbox{\boldmath $\xi$}\biggr]
         +\frac{\partial}{\partial r_k}(B_{0i}B_{0j})\xi_i^{(1)}\frac{\partial\xi_k}{\partial r_j}{\rm div}\mbox{\boldmath $\xi$}.
\label{C.17}
\end{equation}
The first term of equation (\ref{C.17}) is further integrated by parts, neglecting the surface integrals.
Then, equation (\ref{C.17}) is reduced to
\begin{eqnarray}
     &&B_{0i}B_{0j}\biggr[-\frac{1}{2}\frac{\partial\xi_i^{(1)}}{\partial r_j}({\rm div}\mbox{\boldmath $\xi$})^2
          +\xi_i^{(1)}\frac{\partial\xi_k}{\partial r_j}\frac{\partial}{\partial r_k}{\rm div}\mbox{\boldmath $\xi$}\biggr]
                      \nonumber   \\
      &&    -\frac{1}{2}\frac{\partial}{\partial r_j}(B_{0i}B_{0j})\xi_i^{(1)}({\rm div}\mbox{\boldmath $\xi$})^2
         +\frac{\partial}{\partial r_k}(B_{0i}B_{0j})\xi_i^{(1)}\frac{\partial \xi_k}{\partial r_j}{\rm div}\mbox{\boldmath $\xi$}.
\label{C.18}
\end{eqnarray}
Here, we introduce quantities defined by
\begin{equation}
    {\cal A}_{{\rm C}2}(\mbox{\boldmath $\xi$}^{(1)}, \mbox{\boldmath $\xi$}^{(2)}, \mbox{\boldmath $\xi$}^{({\rm D})})
        \equiv -B_{0i}B_{0j}\biggr[\frac{1}{2}\frac{\partial \xi_i^{(1)}}{\partial r_j}({\rm div}\mbox{\boldmath $\xi$})^2
        +{\rm div}\mbox{\boldmath $\xi$}^{(1)}\frac{\partial \xi_i}{\partial r_j}{\rm div}\mbox{\boldmath $\xi$}\biggr].
\label{C.19}
\end{equation}
\begin{equation}
      {\cal A}_{{\rm C}3}(\mbox{\boldmath $\xi$}^{(1)}, \mbox{\boldmath $\xi$}^{(2)}, \mbox{\boldmath $\xi$}^{({\rm D})})
          \equiv -B_{0i}B_{0j}\biggr[\frac{\partial\xi_i^{(1)}}{\partial r_j}\frac{\partial\xi_\ell}{\partial r_k}
                   \frac{\partial\xi_k}{\partial r_\ell}
          +2\frac{\partial\xi_k^{(1)}}{\partial r_\ell}\frac{\partial\xi_i}{\partial r_j}\frac{\partial\xi_\ell}{\partial r_k}\biggr].
\label{C.20}
\end{equation}
Then, equation (\ref{C.16}) can be arranged in the form of 
\begin{eqnarray}
     &&{\cal A}_{{\rm C}2}+{\cal A}_{{\rm C}3}
              \nonumber    \\
      &&+B_{0i}B_{0j}\biggr[\frac{\partial \xi_k^{(1)}}{\partial r_\ell}\frac{\partial\xi_i}{\partial r_j}\frac{\partial\xi_\ell}{\partial r_k}
        +\frac{\partial\xi_i^{(1)}}{\partial r_\ell}\frac{\partial\xi_\ell}{\partial r_k}\frac{\partial\xi_k}{\partial r_j}
        +\xi_i^{(1)}\frac{\partial\xi_k}{\partial r_j}\frac{\partial}{\partial r_k}{\rm div}\mbox{\boldmath $\xi$}\biggr]
                \nonumber   \\
      &&-\frac{1}{2}\frac{\partial}{\partial r_j}(B_{0i}B_{0j})\xi_i^{(1)}({\rm div}\mbox{\boldmath $\xi$})^2
         +\frac{\partial}{\partial r_k}(B_{0i}B_{0j})\xi_i^{(1)}\frac{\partial \xi_k}{\partial r_j}{\rm div}\mbox{\boldmath $\xi$}.
\label{C.21}
\end{eqnarray}

The term of $B_{0i}B_{0j}(\partial\xi_k^{(1)}/\partial r_\ell)(\partial\xi_i/\partial r_j)(\partial\xi_\ell/\partial r_k)$ 
in equation (\ref{C.21}) is integrated 
by parts to lead to $-\xi_i(\partial/\partial r_j)[B_{0i}B_{0j}(\partial\xi_k^{(1)}/\partial r_\ell)(\partial\xi_\ell/\partial r_k)]$.
Furthermore, $B_{0i}B_{0j}(\partial\xi_i^{(1)}/\partial r_\ell)(\partial\xi_\ell/\partial r_k)(\partial\xi_k/\partial r_j)$
is also integrated by part to lead to
$-\xi_i^{(1)}(\partial/\partial r_\ell)[B_{0i}B_{0j}(\partial\xi_\ell/\partial r_k)(\partial \xi_k/\partial r_j)]$.
Then, by introducing ${\cal A}_{{\rm C}4}$ defined by 
\begin{equation}
    {\cal A}_{{\rm C}4}(\mbox{\boldmath $\xi$}^{(1)}, \mbox{\boldmath $\xi$}^{(2)}, \mbox{\boldmath $\xi$}^{({\rm D})})
       \equiv -(B_{0i}B_{0j})\biggr[\xi_i\frac{\partial\xi_k^{(1)}}{\partial r_\ell}\frac{\partial^2\xi_\ell}{\partial r_j\partial r_k}
    +\xi_i\frac{\partial\xi_\ell}{\partial r_k}\frac{\partial^2\xi_k^{(1)}}{\partial r_j\partial r_\ell}
    +\xi_i^{(1)}\frac{\partial\xi_\ell}{\partial r_k}\frac{\partial^2\xi_k}{\partial r_\ell\partial r_j}\biggr],
\label{C.23}
\end{equation}
the sum of the three terms in the second line of equation (\ref{C.21}) is summarized as
\begin{equation}
     {\cal A}_{{\rm C}4}
     -\frac{\partial}{\partial r_j}(B_{0i}B_{0j})\frac{\partial\xi_k^{(1)}}{\partial r_\ell}\xi_i\frac{\partial\xi_\ell}{\partial r_k}
     -\frac{\partial}{\partial r_\ell}(B_{0i}B_{0j})\xi_i^{(1)}\frac{\partial\xi_\ell}{\partial  r_k}\frac{\partial\xi_k}{\partial r_j}.
\label{C.22}
\end{equation}
Hence, equation (\ref{C.16}) is reduced to
\begin{eqnarray}
    &&{\cal A}_{{\rm C}2}+{\cal A}_{{\rm C}3}+{\cal A}_{{\rm C}4}
            \nonumber    \\
    &&-\frac{\partial}{\partial r_j}(B_{0i}B_{0j})\frac{\partial\xi_k^{(1)}}{\partial r_\ell}\xi_i\frac{\partial\xi_\ell}{\partial r_k}
      -\frac{\partial}{\partial r_\ell}(B_{0i}B_{0j})\xi_i^{(1)}\frac{\partial\xi_\ell}{\partial r_k}\frac{\partial\xi_k}{\partial r_j}
            \nonumber    \\
    &&-\frac{1}{2}\frac{\partial}{\partial r_j}(B_{0i}B_{0j})\xi_i^{(1)}({\rm div}\mbox{\boldmath $\xi$})^2
      +\frac{\partial}{\partial r_k}(B_{0i}B_{0j})\xi_i^{(1)}\frac{\partial \xi_k}{\partial r_j}{\rm div}\mbox{\boldmath $\xi$}.
\label{C.24}
\end{eqnarray} 
It should be noticed that in ${\cal A}_{{\rm C}2}$, ${\cal A}_{{\rm C}3}$, and ${\cal A}_{{\rm C}4}$ defined above,
$\mbox{\boldmath $\xi$}^{(1)}$, $\mbox{\boldmath $\xi$}^{(2)}$, and $\mbox{\boldmath $\xi$}^{({\rm D})}$ are all commutative each other,

Summing up the results obtained till the present stage, we can write down $G^{\rm B}$ given by equation (\ref{C.10})
in the form:
\begin{eqnarray}
     G^{\rm B}\Rightarrow &&{\cal A}+{\cal A}_{{\rm C}1} +{\cal A}_{{\rm C}2}+{\cal A}_{{\rm C}3}+{\cal A}_{{\rm C}4}
                +{\cal B}_1+{\cal B}_2+{\cal C}+{\cal D}
                                       \nonumber      \\
     &&-\frac{\partial}{\partial r_j}(B_{0i}B_{0j})\frac{\partial\xi_k^{(1)}}{\partial r_\ell}
                        \xi_i\frac{\partial\xi_\ell}{\partial r_k}
             -\frac{\partial}{\partial r_\ell}(B_{0i}B_{0j})\xi_i^{(1)}\frac{\partial\xi_\ell}{\partial r_k}\frac{\partial\xi_k}{\partial r_j}
                              \nonumber    \\
     &&-\frac{1}{2}\frac{\partial}{\partial r_j}(B_{0i}B_{0j})\xi_i^{(1)}({\rm div}\mbox{\boldmath $\xi$})^2
                      +\frac{\partial}{\partial r_k}(B_{0i}B_{0j})\xi_i^{(1)}
                      \frac{\partial \xi_k}{\partial r_j}{\rm div}\mbox{\boldmath $\xi$}
                               \nonumber     \\
     &&+\frac{\partial\xi_i^{(1)}}{\partial r_j}\Biggr[
               \xi_k{\rm div}\mbox{\boldmath $\xi$}\frac{\partial}{\partial r_k}(B_{0i}B_{0j})
                   +\frac{1}{2}\xi_k\xi_\ell\frac{\partial^2}{\partial r_\ell\partial r_k}(B_{0j}B_{0j})
                   -\frac{\partial\xi_j}{\partial r_k}\xi_\ell\frac{\partial}{\partial r_\ell}(B_{0i}B_{0k})\biggr]
                             \nonumber      \\   
     &&+\frac{1}{2}\xi_i^{(1)}\xi_\ell\xi_k\frac{\partial^3}{\partial r_\ell\partial r_k\partial r_j}
                     (B_{0i}B_{0j})
          +\xi_i^{(1)}\xi_\ell{\rm div}\mbox{\boldmath $\xi$}\frac{\partial^2}{\partial r_\ell\partial r_j}
                     (B_{0i}B_{0j})
                          \nonumber    \\
     &&+\frac{1}{2}\xi_i^{(1)}\biggr[({\rm div}\mbox{\boldmath $\xi$})^2 -\frac{\partial\xi_\ell}{\partial r_k}
                     \frac{\partial \xi_k}{\partial r_\ell}\biggr]
                     \frac{\partial}{\partial r_j}(B_{0i}B_{0j}),
\label{C.25}
\end{eqnarray}
where all the terms below the first line are those with spatial derivative of $B_{0i}B_{0j}$.

The first term on the second line of equation ({\ref{C.25}) and a part of the last line of equation (\ref{C.25})
are summarized in the form of ${\cal B}_{{\rm C}1}$ which is
\begin{equation}
     {\cal B}_{{\rm C}1}(\mbox{\boldmath $\xi$}^{(1)}, \mbox{\boldmath $\xi$}^{(2)}, \mbox{\boldmath $\xi$}^{({\rm D})})
       \equiv -\frac{\partial}{\partial r_j}(B_{0i}B_{0j})\biggr[\frac{1}{2}\xi_i^{(1)}
          \frac{\partial\xi_\ell}{\partial r_k}\frac{\partial\xi_k}{\partial r_\ell}
         +\frac{\partial\xi_k^{(1)}}{\partial r_\ell}\xi_i\frac{\partial\xi_\ell}{\partial r_k}\biggr].
\label{C.26}
\end{equation}

The last term on the third line of equation (\ref{C.25}) and the first term on the fourth line are summarized as 
$(\partial/\partial r_k)(B_{0i}B_{0j})(\partial/\partial r_j)(\xi_i^{(1)}\xi_k){\rm div}\mbox{\boldmath $\xi$}$.
This can be integrated by parts to give
\begin{equation}
   -\frac{\partial^2}{\partial r_k\partial r_j}(B_{0i}B_{0j})\xi_i^{(1)}\xi_k{\rm div}\mbox{\boldmath $\xi$}
   -\frac{\partial}{\partial r_k}(B_{0i}B_{0j})\xi_i^{(1)}\xi_k\frac{\partial}{\partial r_j} {\rm div}\mbox{\boldmath $\xi$}.
\label{C.27}
\end{equation}
The last term is further integrated by parts by writing $(\partial/\partial r_j){\rm div}\mbox{\boldmath $\xi$}$
as $(\partial/\partial_\ell)(\partial\xi_\ell/\partial r_j)$.
Then, equation (\ref{C.27}) is changed as
\begin{equation}
     -\frac{\partial^2}{\partial r_k\partial r_j}(B_{0i}B_{0j})\xi_i^{(1)}\xi_k{\rm div}\mbox{\boldmath $\xi$}
     +\frac{\partial}{\partial r_\ell}\biggr[\frac{\partial}{\partial r_k}(B_{0i}B_{0j})\xi_i^{(1)}\xi_k\biggr]
                 \frac{\partial\xi_\ell}{\partial r_j}.
\label{C.28}
\end{equation}
The first term of equation (\ref{C.28}) and the second term on the fifth line of equation (\ref{C.25}) are cancelled.
The first term on the fifth line of equation (\ref{C.25}) is also integrated by part to give
\begin{equation}
      -\frac{1}{2}\frac{\partial^2}{\partial r_\ell\partial r_k}(B_{0i}B_{0j})
        \biggr[\frac{\partial\xi_i^{(1)}}{\partial r_j}\xi_\ell\xi_k
        +\xi_i^{(1)}\frac{\partial}{\partial r_j}(\xi_\ell\xi_k)\biggr].
\label{C.29}
\end{equation}
The first term of this equation and the second term on the fourth line of equation (\ref{C.25}) are cancelled.

Taking into account the above considerations we find thet the terms below the second line of equation (\ref{C.25})
is summarized as
\begin{eqnarray}
    {\cal B}_{{\rm C}1}
    && -\frac{\partial}{\partial r_\ell}(B_{0i}B_{0j})\xi_i^{(1)}\frac{\partial \xi_\ell}{\partial r_k}\frac{\partial\xi_k}{\partial r_j}
       +\frac{\partial}{\partial r_\ell}\biggr[\frac{\partial}{\partial r_k}(B_{0i}B_{0j})\xi_i^{(1)}\xi_k\biggr]
          \frac{\partial\xi_\ell}{\partial r_j}
             \nonumber     \\
    &&-\frac{\partial\xi_i^{(1)}}{\partial r_j}\frac{\partial\xi_j}{\partial r_k}\xi_\ell\frac{\partial}{\partial r_\ell}(B_{0i}B_{0k})
      -\frac{1}{2}\frac{\partial^2}{\partial r_\ell\partial r_k}(B_{0i}B_{0j})\xi_i^{(1)}\frac{\partial}{\partial r_j}(\xi_\ell\xi_k).
\label{C.30}
\end{eqnarray}    
In the above equation, the terms proportional to $\partial(B_{0i}B_{0j})/\partial r_k$ are cancelled out, and
the coefficient of the term of $\partial^2(B_{0i}B_{0j})/\partial r_\ell\partial r_k$ is 
$(1/2)\xi_i^{(1)}[\xi_k\partial\xi_\ell/\partial r_j-\xi_\ell\partial\xi_k/\partial r_j]$, which vanishes by symmetry.
Consequently, if all the surface integrals are neglected, we have
\begin{eqnarray}
   &&4\pi \int\mbox{\boldmath $\xi$}^{(1)}\cdot\mbox{\boldmath $C$}^{\rm B}(\mbox{\boldmath $\xi$}^{(2)}, \mbox{\boldmath $\xi$}^{({\rm D})})dV
       = \int G^{\rm B}dV      \nonumber  \\
   &&    =\int \biggr({\cal A}+{\cal A}_{{\rm C}1} +{\cal A}_{{\rm C}2}+{\cal A}_{{\rm C}3}+{\cal A}_{{\rm C}4}
             +{\cal B}_1+{\cal B}_2+{\cal B}_{{\rm C}1}+{\cal C}+{\cal D}
              \biggr)dV.
\label{C.31}
\end{eqnarray}
This equation shows the commutative relations:
\begin{equation}
   \int\mbox{\boldmath $\xi$}^{(1)}\cdot\mbox{\boldmath $C$}^{\rm B}(\mbox{\boldmath $\xi$}^{(2)}, \mbox{\boldmath $\xi$}^{({\rm D})})dV
  =\int\mbox{\boldmath $\xi$}^{(2)}\cdot\mbox{\boldmath $C$}^{\rm B}(\mbox{\boldmath $\xi$}^{(1)}, \mbox{\boldmath $\xi$}^{({\rm D})})dV
  =\int\mbox{\boldmath $\xi$}^{({\rm D})}\cdot\mbox{\boldmath $C$}^{\rm B}(\mbox{\boldmath $\xi$}^{(1)}, \mbox{\boldmath $\xi$}^{(2)})dV.
\label{C.32}
\end{equation}

\bigskip\noindent
{\bf Reference}

\leftskip=20pt
\parindent=-20pt
Bernstein, I.B., Frieman, E.A., Kruskal, M.D.,\& Kulsrud, R.M. 1958, Proc. Roy. Soc. (London), A244, 17\par
Cairns, R.A. 1978, J. Fluid Mech., 92,1 \par
Drury, L.O'C. 1985, MNRAS, 217, 821   \par
Fu, W., \& Lai, D. 2009, ApJ, 690, 1386\par
Fu, W., \& Lai, D. 2011, MNRAS, 410, 399 \par
Kato, S. 2001, PASJ, 53, 1\par 
Kato, S. 2004, PASJ, 56, 905\par
Kato, S. 2008, PASJ, 60, 111 \par
Kato, S. 2012, PASJ, 64, 139\par
Kato, S. 2013a, PASJ, 65, 56\par
Kato, S. 2013b, PASJ, 65, 75 (paper I)\par
Kato, S. 2013c, PASJ, submitted  \par
Kato, S., Okazaki, A.-T., \& Oktariani, F. 2011, 63, 365 \par
Khlazov, I.V., Smolyakov, A.I., \& Ilgisonis, V.I., 2008, Phys. of Plasmas, 15, 4501\par
Lai, D. \& Tsang, D. 2009, MNRAS, 393, 979 \par
Lin, C.C., \& Lau, Y.Y., 1980, Studies in Applied Math., 60, 97\par
Lubow, S.H. 1991, ApJ, 381, 259\par
Lynden-Bell, D. and Ostriker, J.P. 1967, MNRAS, 136, 293  \par
Okazaki, A.-T., Kato, S., and Fukue, J. 1987, PASJ, 39, 457 \par
Osaki, Y. 1985, A\&A, 144, 369\par
Papaloizou, J.C.B., \& Pringle, J.E. 1984, MNRAS, 208, 721 \par

\end{document}